\documentclass[aps,pra,twocolumn,floatfix,superscriptaddress,longbibliography]{revtex4-1}

\usepackage{xcolor}
\usepackage{appendix}
\usepackage{graphicx,times}
\usepackage{amstext}
\usepackage{amsmath}            
\usepackage{amssymb}            
\usepackage{graphicx}           
\usepackage{latexsym}
\usepackage{bm}
\usepackage{etoolbox}
\usepackage{breqn}
\usepackage[FIGTOPCAP,raggedright,nooneline]{subfigure}

\makeatletter
\let\cat@comma@active\@empty
\makeatother

\usepackage{url}
\usepackage[colorlinks]{hyperref}
\hypersetup{%
    plainpages=true,
    breaklinks=true,
    hypertexnames=false,
    pageanchor=true,
    colorlinks=true,
    linkcolor={blue},
    citecolor={red},
    urlcolor={blue},
    anchorcolor={black}
}

\newcommand{\beq}{\begin{eqnarray}}
\newcommand{\eeq}{\end{eqnarray}}

\def\<{\langle}
\def\>{\rangle}

\def\<{\langle}
\def\>{\rangle}

\def \info#1{}

\def \info#1{}

\begin{document}

\title{Probing higher-order transitions through scattering of microwave photons in the ultrastrong-coupling regime of circuit QED}

\author{Guan-Ting Chen}
\affiliation{Department of Physics, National Cheng Kung University, Tainan 701, Taiwan}
\author{Po-Chen Kuo}
\affiliation{Department of Physics, National Cheng Kung University, Tainan 701, Taiwan}
\author{Huan-Yu Ku}
\affiliation{Department of Physics, National Cheng Kung University, Tainan 701, Taiwan}
\author{Guang-Yin Chen}
\email{gychen@phys.nchu.edu.tw}
\affiliation{Department of Physics, National Chung Hsing University, Taichung 402, Taiwan}
\author{Yueh-Nan Chen}
\email{yuehnan@mail.ncku.edu.tw}
\affiliation{Department of Physics, National Cheng Kung University, Tainan 701, Taiwan}
\affiliation{Physics Division, National Center for Theoretical Sciences, Hsinchu 300, Taiwan}
\date{\today}

\begin{abstract}
Higher-order transitions can occur in the ultrastrong-coupling regime of circuit QED through virtual processes governed by the counter-rotating interactions. We propose a feasible way to probe higher-order transitions through the scattering of propagating microwave photons incident on the hybrid qubit-cavity system. The lineshapes in the scattering spectra can indicate the coherent interaction between the qubits and the cavity, and the higher-order transitions can be identified in the population spectra. We further find that if the coupling strengths between the two qubits and the cavity are tuned to be asymmetric, the dark antisymmetric state with the Fano-lineshape can also be detected from the variations in the scattering spectra.       
\end{abstract}

\maketitle


\section{INTRODUCTION}

With the recently advanced development in superconducting quantum circuits~(SQCs), investigations of microwave photonics have been extended to circuit quantum electrodynamics~(QED) systems~\cite{You_11,Gu_17,Girvin_09}, in which superconducting artificial atoms and resonators substitute for the essential building blocks (natural atoms and optical cavities) in cavity QED. Superconducting circuit has already been proven as a useful vehicle for the realizations of quantum coherence~\cite{Chiorescu_03}, quantum information processing, and atomic physics~\cite{You_11}, particularly in the regimes not easily accessible with natural atoms and molecules~\cite{Irish_07,Ashhab_10}. In contrast to conventional cavity QED, circuit QED can be artificially designed and fabricated for different research purposes~\cite{Wallraff_04,Hime_06,Niskanen_07,Buluta1_11,Xiang_13,Gu_17}. The energy levels of superconducting artificial atoms and the oscillator frequency of resonator can be adjusted in a wide range of possible values. The coupling strengths between superconducting artificial atoms and their electromagnetic environments can also be tuned. These flexible circuit designs make circuit QED a promising candidate for exploring microwave quantum optics on a superconducting chip.

The interaction between an atomic system and electromagnetic fields in cavity QED has been widely studied over the past few decades~\cite{scully_97,Hood_98,Kimble_98,Mabuchi_02,Hennessy_07}. Analogously, the superconducting artificial atoms in circuit QED can be strongly coupled to quantized microwave fields in the transmission line or 3D resonators~\cite{You_03,Paik_11,Xiang_13,Stern_14,Gu_17}. Just like natural atoms, superconducting artificial atoms are multi-level systems~\cite{You_11}. If we limit our study to the two lowest-energy levels, this can be defined as a superconducting qubit. It is known that the coupling strength~($\lambda$) between the superconducting qubit and the resonator field can be experimentally engineered to become comparable to the transition frequencies of the qubit and the resonator~($\omega_q$ and $\omega_c$, respectively)~\cite{Niemczyk_10,Forn_10,Yoshihara_17,Braumueller_17,Forn_17}. With this extremely strong coupling strength~($\lambda\gtrsim0.1\omega_{q/c}$)~\cite{Gu_17}, one can reach the ultrastrong-coupling~(USC) regime in circuit QED~\cite{Gu_17,Xiang_13,Nataf_10,Peropadre_10,Ridolfo_12,Ridolfo_13,Stassi_13,Garziano_14,Garziano_15,Ma_15,Garziano_16,Anton_17,Forn_17,Niemczyk_10,Forn_10,Yoshihara_17,Braumueller_17}. In the USC regime, higher-order atom-field resonant transitions can occur via virtual processes, which do not conserve the number of excitations\cite{Xiang_13,Gu_17,Stassi_13,Garziano_14,Garziano_15,Ma_15,Garziano_16,Anton_17}. These processes governed by the counter-rotating terms in the interaction Hamiltonian can no longer be neglected, and therefore the rotating wave approximation~(RWA) breaks down~.

Circuit QED is a promising tool to generate single microwave photons~\cite{Houck_07} and also paves the way to study the scattering properties of single microwave photons propagating in the circuit~\cite{Romero_09,Peropadre_11}. Based on this feature, several theoretical works~\cite{Shen_05,Zhou_08,JTShen_09,Witthaut_10,GYChen_14,Roy_17} have been proposed to study the response of injected microwave photons travelling in the transmission line. When a propagating microwave photon is coupled to an emitter, the interaction gives rise to the scattering of the field or the excitation of the emitter. The phenomenon leads to the variations of the profile in the scattering spectra~\cite{Shen_05,Zhou_08,JTShen_09,Witthaut_10,GYChen_14,Roy_17,Chang_07,GYChen_11,WChen_11,GYChen_12,GYChen_16,PCKuo_16}. Therefore, one can use this advantage to measure the qubit states through the detection of the transmitted/reflected photons. While most of the previous studies are focused in the weak- or strong-coupling regimes, in this work, however, we aim to study the scattering spectra of a superconducting circuit system comprising the transmission line resonator coupled to two superconducting charge qubits in the USC regime. The main purpose of this work is to observe the higher-order resonant transitions in populations through the scattering of microwave photons incident on the qubit-cavity system. Moreover, we also find that the dark antisymmetric state in the higher-order transitions can be probed if the coupling strengths between the qubits and the cavity are tuned to be asymmetric. Our consideration provides an experimentally feasible way to detect the higher-order resonant transitions in the USC regime.

\section{THE MODEL}\label{sec:model}

\begin{figure}[th]
\centering
\includegraphics[width=1\columnwidth]{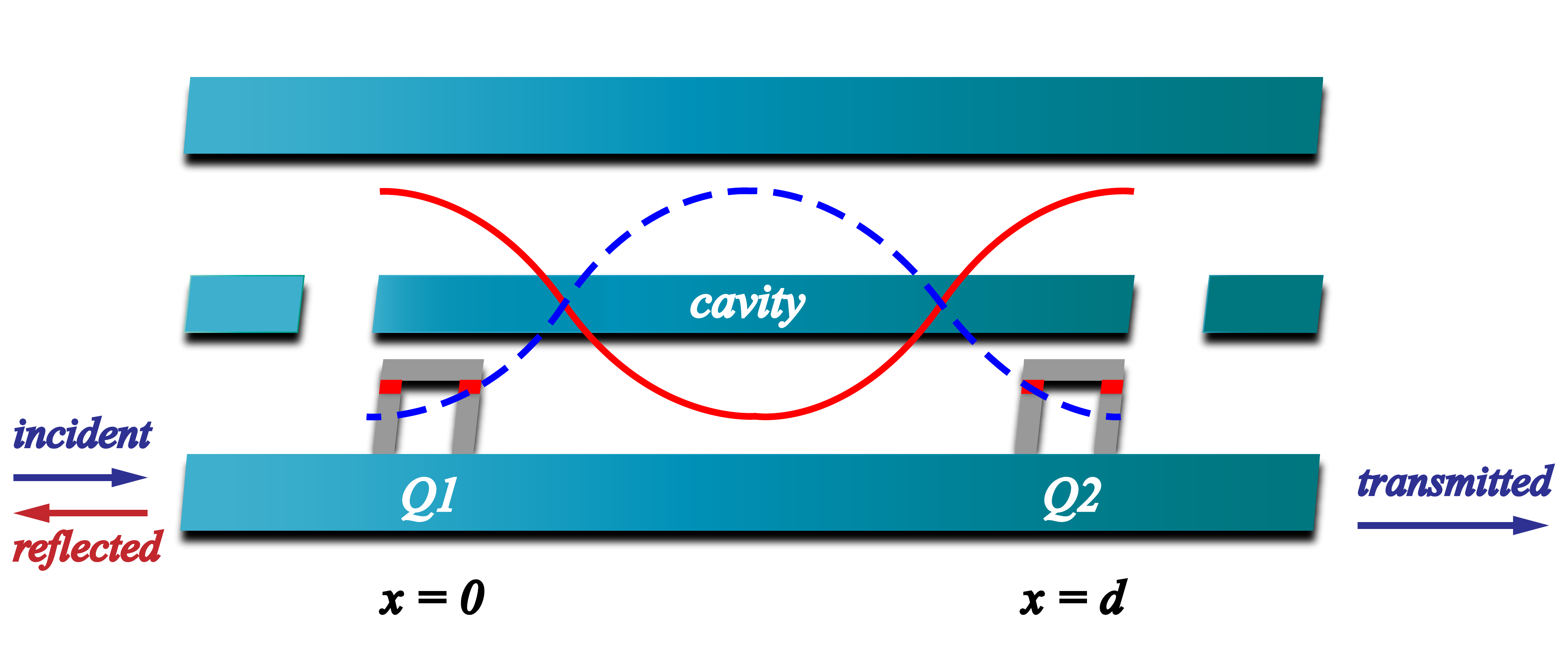}
\caption{(Color online) Schematic diagram of two charge qubits coupled to a resonator in a transmission-line waveguide. A microwave photon incident from the left is coherently scattered by the qubit-cavity system.} \label{fig1}
\end{figure}

We investigate a general one-dimensional model with two identical superconducting qubits, separated by a distance $d$ and embedded in a transmission-line waveguide, coupled to a cavity as depicted in Fig.~\ref{fig1}. We first consider a simple two-level configuration for the two qubits consisting of the ground and the excited states ($|g\rangle, |e\rangle$) with the transition frequency $\omega_q$. The cavity with the resonance frequency $\omega_c$ is formed by two capacitive gaps in the middle conductor. We assume that an incident microwave photon with energy $E_k=\hbar v_gk$ propagating in the transmission line from the left would be either scattered or absorbed by the qubits. Here, $v_g$ and $k$ are the group velocity and wave vector, respectively. The Hamiltonian describing the system can be transformed into real space as $H=H_{wg}+H_I+H_{\textrm{cav}}$, where $H_{\textrm{wg}}$ is the Hamiltonian of the waveguide in which photons propagate, $H_{I}$ describes the interaction between the propagating photons and the two separated identical qubits, and $H_{\textrm{cav}}$ represents the Hamiltonian of a singl-mode microwave cavity field interacting with the two qubits as shown in the following equations~\cite{Garziano_16,GYChen_11}.

\begin{subequations}\label{H:main}
\begin{equation}\label{H:a}
\begin{aligned}
H_{\textrm{wg}}/\hbar&=\int dx(-iv_{g})C_R^{\dag}(x)\frac{\partial}{\partial x}C_R(x)\\
&+\int dx(iv_g)C_L^{\dag}(x)\frac{\partial}{\partial x}C_L(x),\\
\end{aligned}
\end{equation}
\begin{equation}\label{H:b}
\begin{aligned}
H_I/\hbar&=g\int dx\sum_{j=1,2}\delta\Big[x-(j-1)d\Big]\times\\
&\Big[C_R^{\dag}(x)\sigma_-^{(j)}+C_R(x)\sigma_+^{(j)}+C_L^{\dag}(x)\sigma_-^{(j)}+C_L(x)\sigma_+^{(j)}\Big],
\end{aligned}
\end{equation}
\begin{equation}\label{H:c}
\begin{aligned}
H_{\textrm{cav}}/\hbar&=\omega_q\sum_{j=1,2}\sigma_{e_j,e_j}+\omega_ca^{\dag}a\\
&+\sum_{j=1,2}\lambda_j(a+a^\dag)(\cos\theta\sigma_x^{(j)}+\sin\theta\sigma_z^{(j)}).
\end{aligned}
\end{equation}
\end{subequations}

In Eq.~(\ref{H:a}), $C_R^{\dag}(x)\ [C_L^{\dag}(x)]$ denotes a bosonic operator creating a right-going (left-going) photon at x. In Eq.~(\ref{H:b}), $\sigma_+^{(j)}=|e_j\rangle\langle g_j|\ (\sigma_-^{(j)}=|g_j\rangle\langle e_j|)$ represents the raising (lowering) operator of the $jth$ qubit, while $g$ is the coupling strength between the two qubits and the waveguide photon. The first two terms in Eq.~(\ref{H:c}) denote the free Hamiltonians of the qubits and the cavity field with the frequency $\omega_q$ and $\omega_c$. The last term describes the interaction between the $jth$ qubit and the cavity field with the coupling strength $\lambda_j$. The diagonal element of the $jth$ qubit can be represented as the form of $\sigma_{e_j,e_j}=|e_j\rangle\langle e_j|$, and $a^{\dag}$ ($a$) is the creation (annihilation) operator for the cavity photon, while $\sigma_x^{(j)}$ and $\sigma_z^{(j)}$ are Pauli matrices for the $jth$ qubit. One notes that, in this work, we consider the superconducting circuit operating in the USC regime with the qubit-cavity coupling strength $\lambda\geq0.1\omega_q$. The RWA is therefore not valid in the USC regime. Analyzing the properties of the qubits and the cavity in this regime requires the full quantum Rabi Hamiltonian. The Hamiltonian in Eq.~(\ref{H:c}) contains the counter-rotating terms, which do not conserve the number of excitations with the form $\sigma_+^{(j)}a^{\dag}$, $\sigma_-^{(j)}a$, $\sigma_z^{(j)}a^{\dag}$, and $\sigma_z^{(j)}a$. 

The stationary state of the system can be written as
\begin{equation}\label{eq eigenstate}
\begin{aligned}
|E_k\rangle&=\sum_{m=R,L}\sum_{i=1,2}\int dx\phi_{k,m}^\dag(x)C_m^\dag\\
&\times\Big[(1+a^{\dag}+a^{\dag}a^{\dag})(1+\sigma_+^{(i)}+\sigma_+^{(1)}\sigma_+^{(2)})\Big]|\textrm{vac}\rangle\\
&+(1+a^{\dag}+a^{\dag}a^{\dag})\\
&\times\sum_{i=1,2}\sum_{j=0,1,2}\Big[\alpha_{i,j}\sigma_+^{(i)}+\beta_j\sigma_+^{(1)}\sigma_+^{(2)}\gamma_j\Big]|\textrm{vac}\rangle,
\end{aligned}
\end{equation}	
where $|\textrm{vac}\rangle=|g_1,g_2\rangle|0\rangle_{\textrm{cav}}|0\rangle_{\textrm{wg}}$ is the vacuum state with both the superconducting qubits in their ground states and zero photon in both the cavity and the waveguide. Hereafter, we use a simplified notation for the quantum states in this system, for example, $|\textrm{vac}\rangle=|gg00\rangle$. In Eq.~(\ref{eq eigenstate}), $\alpha_{i,j}, \beta_j, and \gamma_j$ are the probability amplitudes of each state: $\alpha_{i,j}$ represents the probability amplitude that the $ith$ qubit absorbs photon energy and jumps to its excited state with $j$ photons existing in the cavity, $\beta_j$ indicates the probability amplitude that both the qubits absorb photon energy and jump to their excited states with $j$ photons existing in the cavity, and $\gamma_j$ denotes the probability amplitude that no qubits absorb photon energy and remain in their ground states with $j$ photons in the cavity. We assume that one photon is incident from the left of the waveguide, the scattering occurs at the position of the two qubits due to their interactions with the incident photon. The scattering amplitude wave function $\phi_{k,R}^\dag(x)$ and $\phi_{k,L}^\dag(x)$ take the form
\begin{equation}\label{eq scattering}
\left\{
\begin{aligned}
\phi_{k,R}^{\dag}(x)&\equiv exp(ikx)[\theta(-x)+a\theta(x)\theta(d-x)+t\theta(x-d)],\\
\phi_{k,L}^{\dag}(x)&\equiv exp(-ikx)[r\theta(-x)+b\theta(x)\theta(d-x)],
\end{aligned}
\right.
\end{equation}     	
where $t$ and $r$ are the transmission and reflection amplitudes, respectively. Here, $a$ and $b$ represent the probability amplitudes of the photon between $x=0$ and $x=d$, while $\theta(x)$ is the unit step function.

In the USC regime, the presence of the counter-rotating terms in the $H_{\textrm{cav}}$ enables four different paths which envolve from the initial state $|gg10\rangle$ to the final state $|ee00\rangle$ via several intermediate virtual states~\cite{Garziano_16}, such as $|ee10\rangle,\ |eg20\rangle,\ |eg10\rangle,\ |eg00\rangle,\ |ge20\rangle, |ge10\rangle,\ |ge00\rangle,\\\ |gg20\rangle,$ and$\ |gg00\rangle$. Without loss of generality, we limit the total Hamiltonian $H$ and the eigenstate $|E_k\rangle$ to the 3-excitation manifold. The Hamiltonian can then be spanned (as shown in Appendix \ref{cal}) by the bases: $\lbrace|ee10\rangle,\ |ee00\rangle,\ |eg20\rangle,\ |eg10\rangle,\ |eg00\rangle,\ |ge20\rangle,\ |ge10\rangle,\\|ge00\rangle,\ |gg20\rangle,\ |gg10\rangle,\ |gg01\rangle_R\ (|gg01\rangle_L),$ and$\ |gg00\rangle\rbrace$. By solving the time-independent eigenvalue equation $H|E_k\rangle=E_k|E_k\rangle$, one can obtain the following relations for the coefficients:
\begin{equation}\label{sol}
\left\{
\begin{aligned}
t&=1+\frac{g}{iv_g}\Big[\alpha_{10}+\alpha_{20}~exp(-ikd)\Big],\\
r&=\frac{g}{iv_g}\Big[\alpha_{10}+\alpha_{20}~exp(ikd)\Big],\\
a&=\frac{g}{iv_g}\alpha_{10}+1,\\
b&=\frac{g}{iv_g}\alpha_{20}~exp(ikd).
\end{aligned}
\right.
\end{equation}
The transmission and reflection amplitudes of the incident microwave photon can then be determined algebraically.

\section{RESULTS AND DISCUSSIONS}\label{sec:result}
\subsection{Higher-order transitions}

\begin{figure}[htbp]
\centering
\includegraphics[width=9cm,height=14cm]{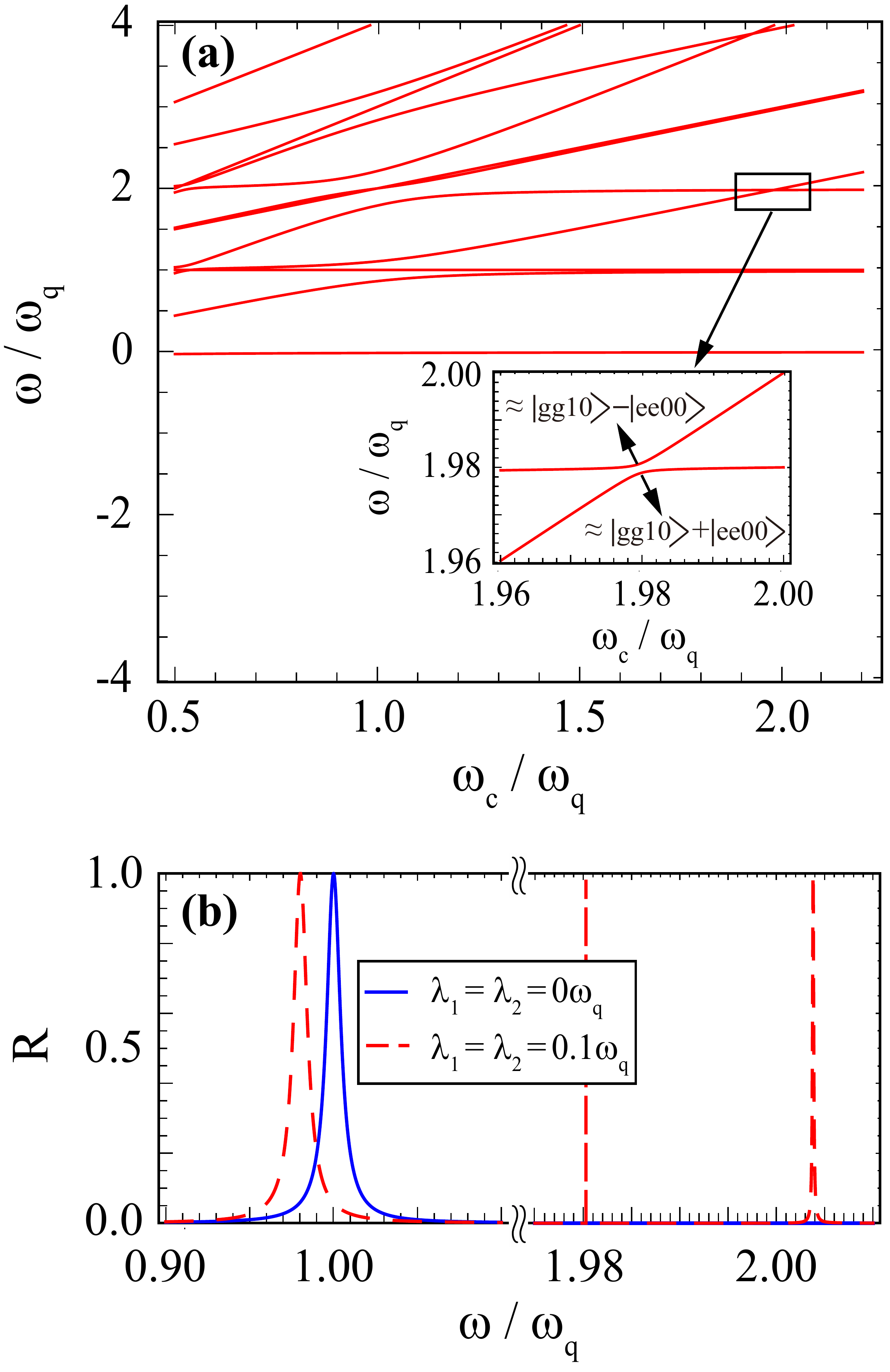}
\caption{(Color online) (a) The figure shows the energy levels for the lowest energy eigenstates of the qubit-cavity system as a function of the normalized cavity frequency $\omega_c/\omega_q$, using the parameters $\lambda_1=\lambda_2=0.1\omega_q$ and $\theta=\pi/6$, where $\omega_q$ is the qubit frequency as a reference point. In the inset, we can observe the anticrossing between $|gg10\rangle$ and $|ee00\rangle$, adopted from~\cite{Garziano_16}. (b) The reflection spectra R of the incident microwave photons in a one-dimensional waveguide for the coupling strengths: $\lambda_1=\lambda_2=0\omega_q$ (blue-solid curve) and $\lambda_1=\lambda_2=0.1\omega_q$ (red-dashed curve).} \label{fig2}
\end{figure}

After numerically diagonalizing the Hamiltonian in Eq.~(\ref{H:c}), the energy levels as a function of the normalized cavity frequency ($\omega_c/\omega_q$) can be plotted as shown in Fig.~2(a). In the region around $\omega_c/\omega_q=2$, a splitting anti-crossing can be observed at energy level $\omega/\omega_q\approx2$ (marked by the black square). It has been reported~\cite{Garziano_16} that the avoided-crossing level [see the inset in Fig.~2(a)] demonstrateing the coupling between the states $|gg10\rangle$ and $|ee00\rangle$ in the USC regime. The interaction does not conserve the number of excitations due to the presence of the counter-rotating terms in the system Hamiltonian. This indicates that if the coupling strength between the qubits and the cavity is sufficiently strong with the frequency of the cavity being double the qubit transition frequency, single photon is able to excite two qubits simultaneously to their excited states even though the cavity is initially in one-photon state~\cite{Garziano_16}.

We now propose a way to feasibly probe the higher-order transitions through the scattering of the microwave photons incident on the hybrid qubit-cavity system. Figure.~2(b) shows the reflection spectra $R=|r|^{2}$ for different coupling strengths $\lambda$ as a function of the normalized microwave photon frequency. We consider the coupling strengths between each qubit and the cavity field, $\lambda_1$ and $\lambda_2$, are the same under the condition of $\omega_c/\omega_q=2$. As can be seen, when the interaction between the qubits and the cavity vanishes, i.e. $\lambda_1=\lambda_2=0$, the peak of the blue-solid curve is on resonance with the qubits. The qubits act like a perfect mirror with total reflection of the incident microwave photons. However, when the couplings are present with $\lambda_1=\lambda_2$, the profile of $R$ (red-dashed curve, with $\lambda_1=\lambda_2=0.1\omega_q$) shows three peaks at the normalized microwave photon frequency ($\omega/\omega_q$) around 0.98, 1.98, and 2.004.

\begin{figure}[htbp]
\centering
\includegraphics[width=9cm,height=7cm]{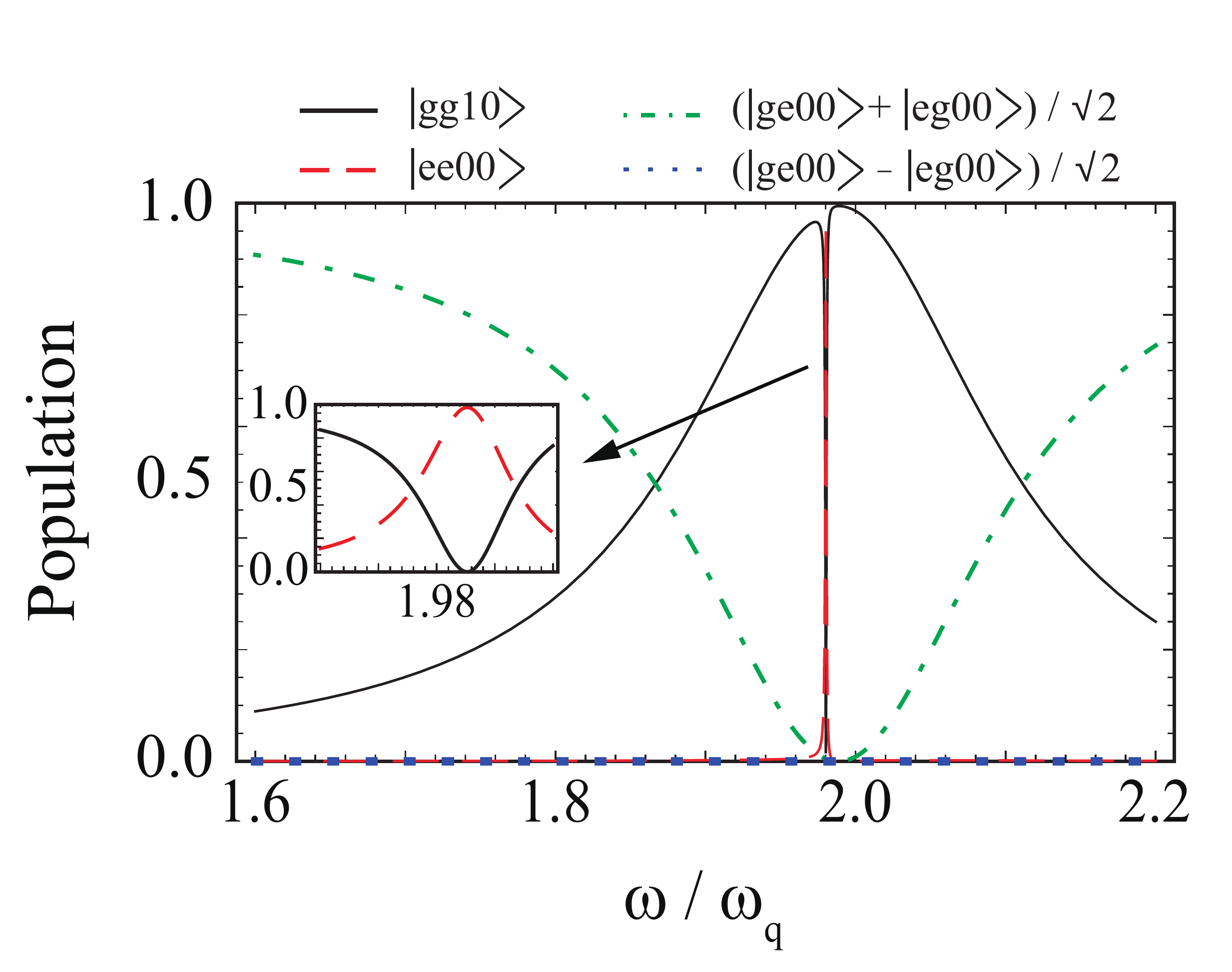}
\caption{(Color online) The populations of the energy eigenstates. The parameters $\lambda_1$, $\lambda_2$, $g$, $\Gamma=\dfrac{4g^{2}}{v^{2}}$ (the decay rate into the microwave photon modes), and $\omega_c$ are normalized to the qubit frequency $\omega_q$. We set $\lambda_1=\lambda_2=0.1\omega_q$, $g=0.05\omega_q$, $\Gamma=0.005\omega_q$, and $\omega_c=2\omega_q$. The black-solid, red-dashed, green-dot-dashed ,and blue-dotted curves represent the states $|gg10\rangle$, $|ee00\rangle$, $(|ge00\rangle+|eg00\rangle)/\sqrt{2}$, and $(|ge00\rangle-|eg00\rangle)/\sqrt{2}$, respectively. In the inset, we can observe the populations of $|ee00\rangle$ and $|gg10\rangle$ exchange at the normalized frequency near 1.98.} \label{fig3}
\end{figure}

Figure.~\ref{fig3} shows the populations of the energy eigenstates as a function of the normalized microwave photon frequency. By comparing Fig.~2(b) with Fig.~\ref{fig3}, one can find that the first peak at the normalized frequency around 0.98 corresponds to the symmetric state \cite{Garziano_16} $(|ge00\rangle+|eg00\rangle)/\sqrt{2}$. When the interaction between the qubits and the cavity exists, the microwave photon can be absorbed by the qubit-cavity system and create one photon in the cavity to form one bare state $|gg10\rangle$. For the frequency of the cavity being double the qubit transition frequency, the bare state $|gg10\rangle$ (black-solid curve) takes all the excitation at the normalized frequency 2.004. Since the initial state $|gg10\rangle$ evolves to the final state $|ee00\rangle$ with the virtual processes \cite{Garziano_16}, we can see that the populations of $|ee00\rangle$ (red-dashed curve) and $|gg10\rangle$ exchange at the normalized frequency near 1.98 in Fig.~\ref{fig3}. Therefore, the second and the third red peaks in Fig.~2(b) can be identified as the states $|ee00\rangle$ and $|gg10\rangle$, respectively. The relation reveals that through the scattering of the microwave photons, one can detect not only the existence of the qubit-cavity interaction, but also the variations of the energy eigenstates by tuning the frequencies of the microwave photons.

\subsection{Dark antisymmetric state}
\begin{figure}[htbp]
\centering
\includegraphics[width=9cm,height=13cm]{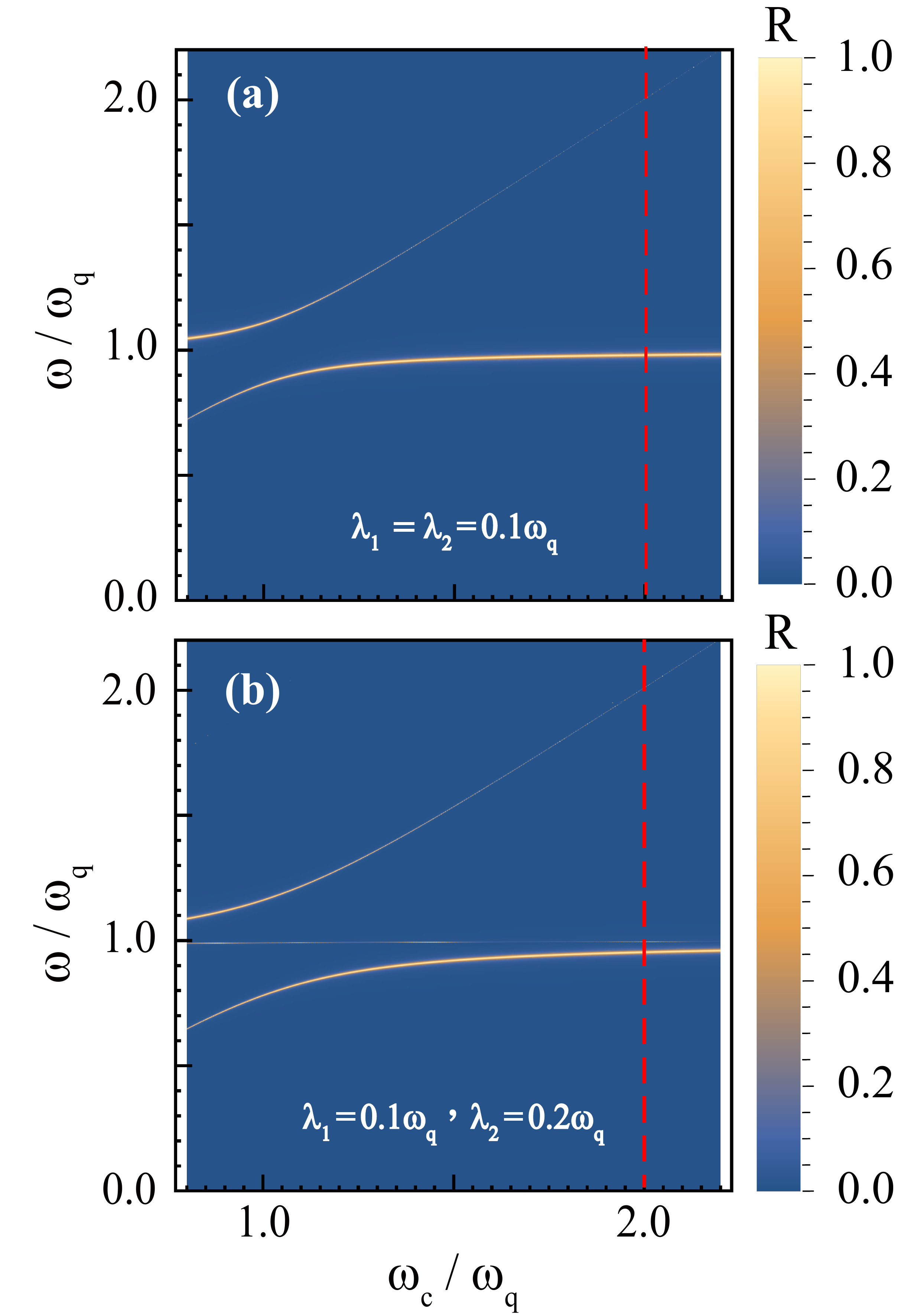}
\caption{(Color online) Figure (a) and (b) are the density plots of the reflection spectra R of the incident microwave photons as functions of $\omega_/\omega_q$ and $\omega_c/\omega_q$. The bright regions indicate the high reflection probability around 1. The coupling strengths between the qubits and the cavity are choosen as $\lambda_1=\lambda_2=0.1\omega_q$ in (a) and $\lambda_1=0.1\omega_q, \lambda_2=0.2\omega_q$ in (b).} \label{fig4}
\end{figure} 

\begin{figure*}[htbp]
\centering
\includegraphics[width=18cm,height=7cm]{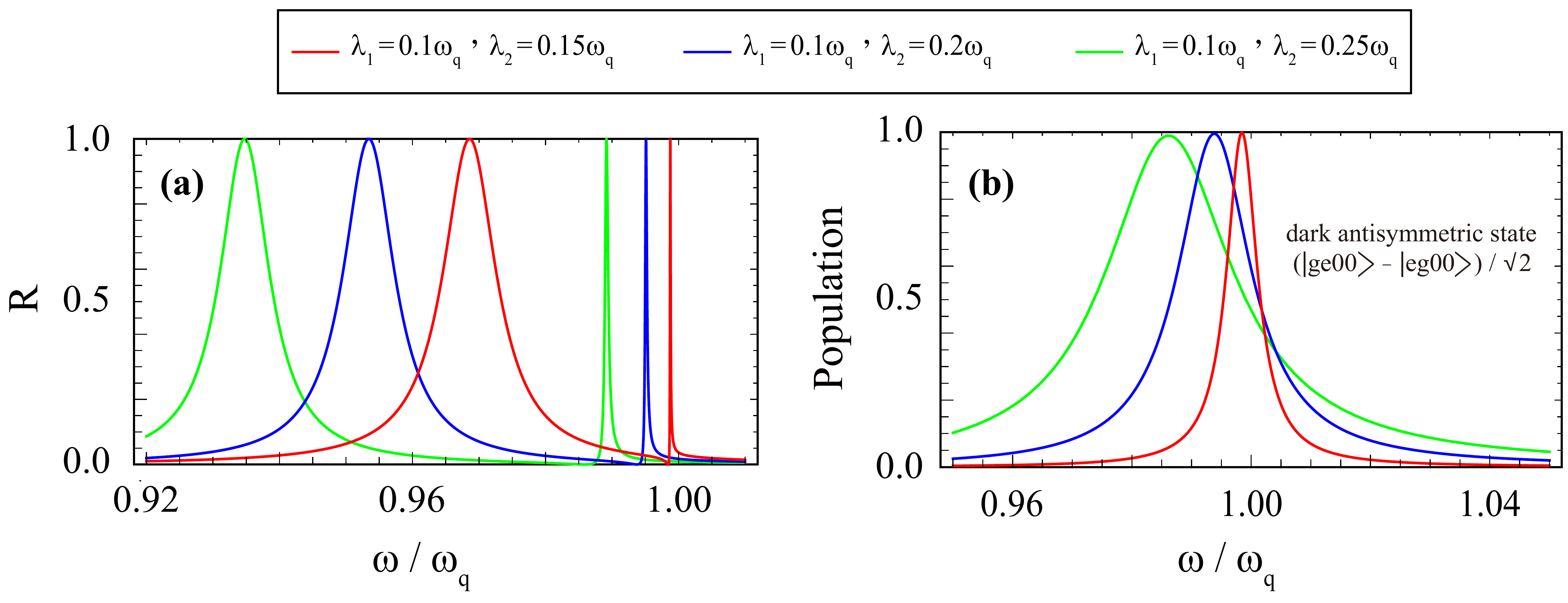}
\caption{(Color online) (a) The reflection spectra R for the asymmetric coupling strengths between the two qubits and the cavity as a function of the normalized microwave photon frequency. (b) The populations of the dark antisymmetric state $(|ge00\rangle-|eg00\rangle)/\sqrt{2}$ as a function of the normalized microwave photon frequency. The red, blue, and green curves denote $(\lambda_1=0.1\omega_q,~\lambda_2=0.15\omega_q)$, $(\lambda_1=0.1\omega_q,~\lambda_2=0.2\omega_q)$, and $(\lambda_1=0.1\omega_q,~\lambda_2=0.25\omega_q)$, respectively.} \label{fig5}
\end{figure*}

In Fig.~2(a), the straight line at $\omega/\omega_q\approx1$ corresponds to the dark antisymmetric state~\cite{Garziano_16} $(|ge00\rangle-|eg00\rangle)/\sqrt{2}$. However, the reflection spectra in the density plot shown in Fig.~4(a) does not exhibit this line for the dark state when $\lambda_1=\lambda_2$. The reason is that the dark state can not absorb/emit photons when $\lambda_1=\lambda_2$. Nevertheless, if the values of $\lambda_1$ and $\lambda_2$ are tuned to be asymmetric, for example, $\lambda_1=0.1\omega_q$ and $\lambda_2=0.2\omega_q$, the line representing the dark state appears at $\omega/\omega_q\approx1$ as shown in Fig.~4(b). This is because the difference between the coupling strengths $\lambda_1$ and $\lambda_2$ changes the eigenvalues, such that the dark state is no longer "dark" and can absorb/emit photons. Along the red-dashed lines placed at $\omega_c/\omega_q=2$ in Figs.~4(a) and 4(b), we can plot the corresponding reflection spectra as a function of the normalized microwave photon frequency as shown in Fig.~2(b) [red-dashed curve] and Fig.~5(a) [blue-solid curve], respectively. In contrast to the red-dashed curve ($\lambda_1=\lambda_2=0.1\omega_q$) in Fig.~2(b), each curve in Fig.~5(a) shows an extra Fano-like lineshape located at $\omega/\omega_q\approx1$ due to the difference between $\lambda_1$ and $\lambda_2$. The Fano lineshapes are more distinct when $\lambda_2$ is close to $\lambda_1$, and the peaks of the Fano lineshapes shift to left as the values of $\lambda_2$ increase. Figure~5(b) displays the populations of the dark antisymmetric states [$(|ge00\rangle-|eg00\rangle)/\sqrt{2}$] for different asymmetric coupling strengths between the two qubits and the cavity. Here, the value of $\lambda_1$ is fixed at $0.1\omega_q$, and the values of $\lambda_2$ are chosen as $1.5\omega_q$, $2.0\omega_q$, and $2.5\omega_q$ (red, blue, and green curves, respectively). When the values of $\lambda_2$ are not equal to $\lambda_1$, one can find that the dark antisymmetric states show the Lorentzian lineshapes and the populations can reach unity. This is totally different from the zero popultion of the truely dark antisymmetric state (blue-dotted curve) in Fig.~\ref{fig3}. The peaks of the Lorentzian lineshapes also shift to left as the values of $\lambda_2$ increase. The trend coincides with the Fano lineshapes in Fig.~5(a). The Fano resonance can occur when the localized channel (discrete state) interfere with the the delocalized channel (continuum)~\cite{Miroshnichenko_10, GYChen_17, PCKuo_16}. Here, the Lorentzian lineshapes of the dark antisymmetric states in Fig.~5(b) represent the localized channel, and the microwave photons stand for the delocalized channel. The curves in Fig.~5(a) therefore exhibit asymmetric lineshapes of the Fano resonance at $\omega/\omega_q\approx1$.

The above results show that, by analyzing the scattering spectra, one can probe the higher-order resonant transitions stemming from the interplay of the energy levels in the USC regime. Experimentally, the coupling strengths between the qubits and the microwave photons can be tuned by changing the magnetic flux and the gate voltage, and the detuning between the transition frequency of the qubit and the microwave photons can be also changed in a similar way.

\section{CONCLUSION}\label{sec:conclusion}
In conclusion, we investigate the superconducting circuit system consisting of two charged qubits coupled to the cavity in a transmission-line waveguide in the USC regime. We propose that through the scattering of the microwave photons incident on the qubit-cavity system, one can probe higher-order qubit-cavity resonant transitions which do not conserve the number of excitations and cannot be observed in the weak- and strong-coupling regimes. We further show that by tuning the two coupling strengths $\lambda_1$ and $\lambda_2$ properly, the dark antisymmetric state with the Fano lineshape can be also detected in the scattering spectra. Our proposal provides an experimentally feasible way to observe the interesting phenomena in the USC regime.

\section*{ACKNOWLEDGMENTS}\label{sec:acknowledgments}
This work is supported partially by the National Center for Theoretical Sciences and Ministry of Science and Technology, Taiwan, under Grant No. MOST 103-2112-M-006-017-MY4 and MOST 105-2112-M-005-008-MY3.  

\begin{widetext}
\appendix
\section{Analytical calculations of the transmission and reflection amplitudes}\label{cal}

In this appendix, we present the full calculations for solving the time-independent eigenvalue equation $H|E_k\rangle=E_k|E_k\rangle$. First, we represent the Hamiltonian in  Eq.~(\ref{H:main}) with the matrix form as shown below. In Eq.~(\ref{trun:main}), $H_{R}$ describes the total Hamiltonian with an incident microwave photon propagating in right direction, while $H_{L}$ describes the Hamiltonian that a microwave photon propagates to left direction. The bases from left to right in both $H_{R}$ and $H_{L}$ are $|ee10\rangle,\ |ee00\rangle,\ |eg20\rangle,\ |eg10\rangle,\ |eg00\rangle,\ |ge20\rangle,\ |ge10\rangle,\ |ge00\rangle,\ |gg20\rangle,\ |gg10\rangle,\ |gg01\rangle_{R/L},$ and$\ |gg00\rangle$. 

\begin{subequations}\label{trun:main}  
\begin{equation}\label{trun:R}
H_{R}/\hbar =
\left(
\begin{array}{cccccccccccc}
C_4 &C_9  &\sqrt{2}C_8 &0 &C_8 &\sqrt{2} C_7 &0 &C_7 &0 &0 &0 &0 \\
C_9 &2C_2 &0 &C_8 &0 &0 &C_7 &0 &0 &0 &0 &0 \\
\sqrt{2} C_8 &0 &C_5 &\sqrt{2} C_{10} &0 &0 &0 &0 &0 &\sqrt{2} C_7 &0 &0 \\
0 &C_8 &\sqrt{2} C_{10} &C_6 &C_{10} &0 &0 &0 &\sqrt{2} C_7 &0 &0 &C_7 \\
C_8 &0 &0 &C_{10} &C_2 &0 &0 &0 &0 &C_7 &k_2 &0 \\
\sqrt{2} C_7 &0 &0 &0 &0 &C_5 &-\sqrt{2} C_{10} &0 &0 &\sqrt{2} C_8 &0 &0 \\
0 &C_7 &0 &0 &0 &-\sqrt{2} C_{10} &C_6 &-C_{10} &\sqrt{2} C_8 &0 &0 &C_8 \\
C_7 &0 &0 &0 &0 &0 &-C_{10} &C_2 &0 &C_8 &k_2 &0 \\
0 &0 &0 &\sqrt{2} C_7 &0 &0 &\sqrt{2} C_8 &0 &2C_3 &-\sqrt{2} C_9 &0 &0 \\
0 &0 &\sqrt{2} C_7 &0 &C_7 &\sqrt{2} C_8 &0 &C_8 &-\sqrt{2} C_9 &C_3 &0 &-C_9 \\
0 &0 &0 &0 &k_2 &0 &0 &k_2 &0 &0 &C_1k_1 &0 \\
0 &0 &0 &C_7 &0 &0 &C_8 &0 &0 &-C_9 &0 &0 
\end{array}
\right),
\end{equation}
\begin{equation}\label{trun:L}
H_{L}/\hbar =
\left(
\begin{array}{cccccccccccc}
0 &0 &0 &0 &0 &0 &0 &0 &0 &0 &0 &0 \\
0 &0 &0 &0 &0 &0 &0 &0 &0 &0 &0 &0 \\
0 &0 &0 &0 &0 &0 &0 &0 &0 &0 &0 &0 \\
0 &0 &0 &0 &0 &0 &0 &0 &0 &0 &0 &0 \\
0 &0 &0 &0 &0 &0 &0 &0 &0 &0 &k_2 &0 \\
0 &0 &0 &0 &0 &0 &0 &0 &0 &0 &0 &0 \\
0 &0 &0 &0 &0 &0 &0 &0 &0 &0 &0 &0 \\
0 &0 &0 &0 &0 &0 &0 &0 &0 &0 &k_3 &0 \\
0 &0 &0 &0 &0 &0 &0 &0 &0 &0 &0 &0 \\
0 &0 &0 &0 &0 &0 &0 &0 &0 &0 &0 &0 \\
0 &0 &0 &0 &k_2 &0 &0 &k_3 &0 &0 &-C_1k_1 &0 \\
0 &0 &0 &0 &0 &0 &0 &0 &0 &0 &0 &0 \\
\end{array}
\right),
\end{equation}
\end{subequations}
with the values $C_1$ to $C_{10}$ and $k_1$ to $k_3$ defined as
\[
\left\{
\begin{array}{lll}
C_1=-i\Gamma\dfrac{v^{2}}{4g^{2}}&, C_2=1&, C_3=\delta,\\
C_4=2+\delta&, C_5=1+2\delta&, C_6=1+\delta,\\
C_7=\lambda_a\cos\theta&, C_8=\lambda_b\cos\theta&, C_9=(\lambda_a+\lambda_b)\sin\theta,\\
C_{10}=(\lambda_a-\lambda_b)\sin\theta,\\
k_1=\int\frac{\partial}{\partial x}dx&, k_2=\int g\delta(x)dx&, k_3=\int g\delta(x-d)dx.
\end{array}
\right.
\]

Note that $C_1$ to $C_{10}$ are all normalized to the qubit frequency $\omega_q$, and therefore $\lambda_a$, $\lambda_b$, and $\delta$ represent $\lambda_1/\omega_q$, $\lambda_2/\omega_q$, and $\omega_c/\omega_q$, respectively.

The stationary eigenstates of the system can be written with the matrix form:
\begin{equation}\label{trun_EkR}
|E_k\rangle_R =
\left(
\begin{array}{cccccccccccc}
\beta_1 \\ \beta_0 \\ \sqrt{2} \alpha_{12} \\ \alpha_{11} \\ \alpha_{10} \\ \sqrt{2} \alpha_{22} \\ \alpha_{21} \\ \alpha_{20} \\ \sqrt{2} \gamma_2 \\ \gamma_1 \\ \phi_{k,R}^{\dag}(x) \\ \gamma_0 
\end{array}
\right),
|E_k\rangle_L =
\left(
\begin{array}{cccccccccccc}
\beta_1 \\ \beta_0 \\ \sqrt{2} \alpha_{12} \\ \alpha_{11} \\ \alpha_{10} \\ \sqrt{2} \alpha_{22} \\ \alpha_{21} \\ \alpha_{20} \\ \sqrt{2} \gamma_2 \\ \gamma_1 \\ \phi_{k,L}^{\dag}(x) \\ \gamma_0 
\end{array}
\right).
\end{equation}

By solving the time-independent Schr\"{o}dinger equation, $H_R|E_k\rangle_R=E_k|E_k\rangle_R$ and $H_L|E_k\rangle_L=E_k|E_k\rangle_L$, one obtains the following equations:  
\begin{equation}\label{eq1}
C_8\alpha_{10}+2C_8\alpha_{12}+C_7\alpha_{20}+2C_7\alpha_{22}+C_9\beta_0+(C_4-v_gk)\beta_1=0,
\end{equation}
\begin{equation}\label{eq2}
C_8\alpha_{11}+C_7\alpha_{21}+(2C_2-v_gk)\beta_0+C_9\beta_1=0,
\end{equation}
\begin{equation}\label{eq3}
\sqrt{2}C_{10}\alpha_{11}+(\sqrt{2}C_5-v_gk)\alpha_{12}+\sqrt{2}C_8\beta_1+\sqrt{2}C_7\gamma_1=0,
\end{equation}
\begin{equation}\label{eq4}
C_{10}\alpha_{10}+(C_6-v_gk)\alpha_{11}+2C_{10}\alpha_{12}+C_8\beta_0+C_7\gamma_0+2C_7\gamma_2=0,
\end{equation}
\begin{equation}\label{eq5}
(C_2-v_gk)\alpha_{10}+C_{10}\alpha_{11}+C_8\beta_1+C_7\gamma_1=-k_2\int dx[\phi_{k,R}^{\dag}(x)+\phi_{k,L}^{\dag}(x)],
\end{equation}
\begin{equation}\label{eq6}
-\sqrt{2}C_{10}\alpha_{21}+(\sqrt{2}C_5-v_gk)\alpha_{22}+\sqrt{2}C_7\beta_1+\sqrt{2}C_8\gamma_1=0,
\end{equation}
\begin{equation}\label{eq7}
-C_{10}\alpha_{20}+(C_6-v_gk)\alpha_{21}+-2C_{10}\alpha_{22}+C_7\beta_0+C_8\gamma_0+2C_8\gamma_2=0,
\end{equation}
\begin{equation}\label{eq8}
(C_2-v_gk)\alpha_{20}-C_{10}\alpha_{21}+C_7\beta_1+C_8\gamma_1=-k_3\int dx[\phi_{k,R}^{\dag}(x)+\phi_{k,L}^{\dag}(x)],
\end{equation}
\begin{equation}\label{eq9}
\sqrt{2}C_7\alpha_{11}+\sqrt{2}C_8\alpha_{21}-\sqrt{2}C_9\gamma_1+(2\sqrt{2}C_3-v_gk)\gamma_2=0,
\end{equation}
\begin{equation}\label{eq10}
C_7\alpha_{10}+2C_7\alpha_{12}+C_8\alpha_{20}+2C_8\alpha_{22}-C_9\gamma_0+(C_3-v_gk)\gamma_1-2C_9\gamma_2=0,
\end{equation}
\begin{equation}\label{eq11}
C_7\alpha_{11}+C_8\alpha_{21}-v_gk\gamma_0-C_9\gamma_1=0,
\end{equation}
\begin{subequations}\label{eq12:main}
\begin{equation}\label{eq12:a}
C_1k_1\phi_{k,R}^{\dag}(x)+k_2\alpha_{10}+k_3\alpha_{20}=v_gk\int dx\phi_{k,R}^{\dag}(x),
\end{equation}
\begin{equation}\label{eq12:b}
-C_1k_1\phi_{k,L}^{\dag}(x)+k_2\alpha_{10}+k_3\alpha_{20}=v_gk\int dx\phi_{k,L}^{\dag}(x).
\end{equation}
\end{subequations}

Our goal is to obtain the transmission and reflection amplitudes for the incident microwave photons. We then compute Eqs.~(\ref{eq12:main}) with $\phi_{k,R}^{\dag}(x)=exp(ikx)[\theta(-x)+a\theta(x)\theta(d-x)+t\theta(x-d)]$ and $\phi_{k,L}^{\dag}(x)=exp(-ikx)[r\theta(-x)+b\theta(x)\theta(d-x)]$ to have the following coefficients:

\begin{subequations}\label{eq13:main}
\begin{equation}\label{eq13:a}
t=1+\frac{g}{iv_g}\Big[\alpha_{10}+\alpha_{20}~exp(-ikd)\Big],
\end{equation}
\begin{equation}\label{eq13:b}
r=\frac{g}{iv_g}\Big[\alpha_{10}+\alpha_{20}~exp(ikd)\Big],
\end{equation}
\begin{equation}\label{eq13:c}
a=\frac{g}{iv_g}\alpha_{10}+1,
\end{equation}
\begin{equation}\label{eq13:d}
b=\frac{g}{iv_g}\alpha_{20}~exp(ikd).
\end{equation}
\end{subequations}

Finally, we obtain the solutions of $\alpha_{10}$ and $\alpha_{20}$ of Eqs.~(\ref{eq1})-(\ref{eq11}). By further substituting the solutions of $\alpha_{10}$ and $\alpha_{20}$ into Eqs.~(\ref{eq13:a}) and (\ref{eq13:b}), one can obtain the transmission and reflection amplitudes. 
\end{widetext}  

%

\begin{thebibliography}{51}%
\makeatletter
\providecommand \@ifxundefined [1]{%
 \@ifx{#1\undefined}
}%
\providecommand \@ifnum [1]{%
 \ifnum #1\expandafter \@firstoftwo
 \else \expandafter \@secondoftwo
 \fi
}%
\providecommand \@ifx [1]{%
 \ifx #1\expandafter \@firstoftwo
 \else \expandafter \@secondoftwo
 \fi
}%
\providecommand \natexlab [1]{#1}%
\providecommand \enquote  [1]{``#1''}%
\providecommand \bibnamefont  [1]{#1}%
\providecommand \bibfnamefont [1]{#1}%
\providecommand \citenamefont [1]{#1}%
\providecommand \href@noop [0]{\@secondoftwo}%
\providecommand \href [0]{\begingroup \@sanitize@url \@href}%
\providecommand \@href[1]{\@@startlink{#1}\@@href}%
\providecommand \@@href[1]{\endgroup#1\@@endlink}%
\providecommand \@sanitize@url [0]{\catcode `\\12\catcode `\$12\catcode
  `\&12\catcode `\#12\catcode `\^12\catcode `\_12\catcode `\%12\relax}%
\providecommand \@@startlink[1]{}%
\providecommand \@@endlink[0]{}%
\providecommand \url  [0]{\begingroup\@sanitize@url \@url }%
\providecommand \@url [1]{\endgroup\@href {#1}{\urlprefix }}%
\providecommand \urlprefix  [0]{URL }%
\providecommand \Eprint [0]{\href }%
\providecommand \doibase [0]{http://dx.doi.org/}%
\providecommand \selectlanguage [0]{\@gobble}%
\providecommand \bibinfo  [0]{\@secondoftwo}%
\providecommand \bibfield  [0]{\@secondoftwo}%
\providecommand \translation [1]{[#1]}%
\providecommand \BibitemOpen [0]{}%
\providecommand \bibitemStop [0]{}%
\providecommand \bibitemNoStop [0]{.\EOS\space}%
\providecommand \EOS [0]{\spacefactor3000\relax}%
\providecommand \BibitemShut  [1]{\csname bibitem#1\endcsname}%
\let\auto@bib@innerbib\@empty
\bibitem [{\citenamefont {You}\ and\ \citenamefont {Nori}(2011)}]{You_11}%
  \BibitemOpen
  \bibfield  {author} {\bibinfo {author} {\bibfnamefont {J.~Q.}\ \bibnamefont
  {You}}\ and\ \bibinfo {author} {\bibfnamefont {F.}~\bibnamefont {Nori}},\
  }\bibfield  {title} {\enquote {\bibinfo {title} {Atomic physics and quantum
  optics using superconducting circuits},}\ }\href {\doibase
  10.1038/nature10122} {\bibfield  {journal} {\bibinfo  {journal} {Nature}\
  }\textbf {\bibinfo {volume} {474}},\ \bibinfo {pages} {589} (\bibinfo {year}
  {2011})}\BibitemShut {NoStop}%
\bibitem [{\citenamefont {Gu}\ \emph {et~al.}(2017)\citenamefont {Gu},
  \citenamefont {Kockum}, \citenamefont {Miranowicz}, \citenamefont {Liu},\
  and\ \citenamefont {Nori}}]{Gu_17}%
  \BibitemOpen
  \bibfield  {author} {\bibinfo {author} {\bibfnamefont {X.}~\bibnamefont
  {Gu}}, \bibinfo {author} {\bibfnamefont {A.~F.}\ \bibnamefont {Kockum}},
  \bibinfo {author} {\bibfnamefont {A.}~\bibnamefont {Miranowicz}}, \bibinfo
  {author} {\bibfnamefont {Y.~X.}\ \bibnamefont {Liu}}, \ and\ \bibinfo
  {author} {\bibfnamefont {F.}~\bibnamefont {Nori}},\ }\bibfield  {title}
  {\enquote {\bibinfo {title} {Microwave photonics with superconducting quantum
  circuits},}\ }\href {\doibase 10.1016/j.physrep.2017.10.002} {\bibfield
  {journal} {\bibinfo  {journal} {Phys. Rep.}\ }\textbf {\bibinfo {volume}
  {718-719}} (\bibinfo {year} {2017}),\
  10.1016/j.physrep.2017.10.002}\BibitemShut {NoStop}%
\bibitem [{\citenamefont {Girvin}(2014)}]{Girvin_09}%
  \BibitemOpen
  \bibfield  {author} {\bibinfo {author} {\bibfnamefont {S.~M.}\ \bibnamefont
  {Girvin}},\ }\href@noop {} {\enquote {\bibinfo {title} {Circuit {QED}:
  superconducting qubits coupled to microwave photons},}\ } (\bibinfo {year}
  {2014})\BibitemShut {NoStop}%
\bibitem [{\citenamefont {Chiorescu}\ \emph {et~al.}(2003)\citenamefont
  {Chiorescu}, \citenamefont {Nakamura}, \citenamefont {Harmans},\ and\
  \citenamefont {Mooij}}]{Chiorescu_03}%
  \BibitemOpen
  \bibfield  {author} {\bibinfo {author} {\bibfnamefont {I.}~\bibnamefont
  {Chiorescu}}, \bibinfo {author} {\bibfnamefont {Y.}~\bibnamefont {Nakamura}},
  \bibinfo {author} {\bibfnamefont {C.~J.~P.~M.}\ \bibnamefont {Harmans}}, \
  and\ \bibinfo {author} {\bibfnamefont {J.~E.}\ \bibnamefont {Mooij}},\
  }\bibfield  {title} {\enquote {\bibinfo {title} {Coherent quantum dynamics of
  a superconducting flux qubit},}\ }\href {\doibase 10.1126/science.1081045}
  {\bibfield  {journal} {\bibinfo  {journal} {Science}\ }\textbf {\bibinfo
  {volume} {299}},\ \bibinfo {pages} {1869} (\bibinfo {year}
  {2003})}\BibitemShut {NoStop}%
\bibitem [{\citenamefont {Irish}(2007)}]{Irish_07}%
  \BibitemOpen
  \bibfield  {author} {\bibinfo {author} {\bibfnamefont {E.~K.}\ \bibnamefont
  {Irish}},\ }\bibfield  {title} {\enquote {\bibinfo {title} {Generalized
  rotating-wave approximation for arbitrarily large coupling},}\ }\href
  {\doibase 10.1103/PhysRevLett.99.173601} {\bibfield  {journal} {\bibinfo
  {journal} {Phys. Rev. Lett.}\ }\textbf {\bibinfo {volume} {99}},\ \bibinfo
  {pages} {173601} (\bibinfo {year} {2007})}\BibitemShut {NoStop}%
\bibitem [{\citenamefont {Ashhab}\ and\ \citenamefont
  {Nori}(2010)}]{Ashhab_10}%
  \BibitemOpen
  \bibfield  {author} {\bibinfo {author} {\bibfnamefont {S.}~\bibnamefont
  {Ashhab}}\ and\ \bibinfo {author} {\bibfnamefont {F.}~\bibnamefont {Nori}},\
  }\bibfield  {title} {\enquote {\bibinfo {title} {Qubit-oscillator systems in
  the ultrastrong-coupling regime and their potential for preparing
  nonclassical states},}\ }\href {\doibase 10.1103/PhysRevA.81.042311}
  {\bibfield  {journal} {\bibinfo  {journal} {Phys. Rev. A}\ }\textbf {\bibinfo
  {volume} {81}},\ \bibinfo {pages} {042311} (\bibinfo {year}
  {2010})}\BibitemShut {NoStop}%
\bibitem [{\citenamefont {Wallraff}\ \emph {et~al.}(2004)\citenamefont
  {Wallraff}, \citenamefont {Schuster}, \citenamefont {Blais}, \citenamefont
  {Frunzio}, \citenamefont {Huang}, \citenamefont {Majer}, \citenamefont
  {Kumar}, \citenamefont {Girvin},\ and\ \citenamefont
  {Schoelkopf}}]{Wallraff_04}%
  \BibitemOpen
  \bibfield  {author} {\bibinfo {author} {\bibfnamefont {A.}~\bibnamefont
  {Wallraff}}, \bibinfo {author} {\bibfnamefont {D.~I.}\ \bibnamefont
  {Schuster}}, \bibinfo {author} {\bibfnamefont {A.}~\bibnamefont {Blais}},
  \bibinfo {author} {\bibfnamefont {L.}~\bibnamefont {Frunzio}}, \bibinfo
  {author} {\bibfnamefont {R.~S.}\ \bibnamefont {Huang}}, \bibinfo {author}
  {\bibfnamefont {J.}~\bibnamefont {Majer}}, \bibinfo {author} {\bibfnamefont
  {S.}~\bibnamefont {Kumar}}, \bibinfo {author} {\bibfnamefont {S.~M.}\
  \bibnamefont {Girvin}}, \ and\ \bibinfo {author} {\bibfnamefont {R.~J.}\
  \bibnamefont {Schoelkopf}},\ }\bibfield  {title} {\enquote {\bibinfo {title}
  {Strong coupling of a single photon to a superconducting qubit using circuit
  quantum electrodynamics},}\ }\href {\doibase 10.1038/nature02851} {\bibfield
  {journal} {\bibinfo  {journal} {Nature}\ }\textbf {\bibinfo {volume} {431}},\
  \bibinfo {pages} {162} (\bibinfo {year} {2004})}\BibitemShut {NoStop}%
\bibitem [{\citenamefont {Hime}\ \emph {et~al.}(2006)\citenamefont {Hime},
  \citenamefont {Reichardt}, \citenamefont {Plourde}, \citenamefont
  {Robertson}, \citenamefont {Wu}, \citenamefont {Ustinov},\ and\ \citenamefont
  {Clarke}}]{Hime_06}%
  \BibitemOpen
  \bibfield  {author} {\bibinfo {author} {\bibfnamefont {T.}~\bibnamefont
  {Hime}}, \bibinfo {author} {\bibfnamefont {P.~A.}\ \bibnamefont {Reichardt}},
  \bibinfo {author} {\bibfnamefont {B.~L.~T.}\ \bibnamefont {Plourde}},
  \bibinfo {author} {\bibfnamefont {T.~L.}\ \bibnamefont {Robertson}}, \bibinfo
  {author} {\bibfnamefont {C.~E.}\ \bibnamefont {Wu}}, \bibinfo {author}
  {\bibfnamefont {A.~V.}\ \bibnamefont {Ustinov}}, \ and\ \bibinfo {author}
  {\bibfnamefont {J.}~\bibnamefont {Clarke}},\ }\bibfield  {title} {\enquote
  {\bibinfo {title} {Solid-state qubits with current-controlled coupling},}\
  }\href {\doibase 10.1126/science.1134388} {\bibfield  {journal} {\bibinfo
  {journal} {Science}\ }\textbf {\bibinfo {volume} {314}},\ \bibinfo {pages}
  {1427} (\bibinfo {year} {2006})}\BibitemShut {NoStop}%
\bibitem [{\citenamefont {Niskanen}\ \emph {et~al.}(2007)\citenamefont
  {Niskanen}, \citenamefont {Harrabi}, \citenamefont {Yoshihara}, \citenamefont
  {Nakamura}, \citenamefont {Lloyd},\ and\ \citenamefont {Tsai}}]{Niskanen_07}%
  \BibitemOpen
  \bibfield  {author} {\bibinfo {author} {\bibfnamefont {A.~O.}\ \bibnamefont
  {Niskanen}}, \bibinfo {author} {\bibfnamefont {K.}~\bibnamefont {Harrabi}},
  \bibinfo {author} {\bibfnamefont {F.}~\bibnamefont {Yoshihara}}, \bibinfo
  {author} {\bibfnamefont {Y.}~\bibnamefont {Nakamura}}, \bibinfo {author}
  {\bibfnamefont {S.}~\bibnamefont {Lloyd}}, \ and\ \bibinfo {author}
  {\bibfnamefont {J.~S.}\ \bibnamefont {Tsai}},\ }\bibfield  {title} {\enquote
  {\bibinfo {title} {Quantum coherent tunable coupling of superconducting
  qubits},}\ }\href {\doibase 10.1126/science.1141324} {\bibfield  {journal}
  {\bibinfo  {journal} {Science}\ }\textbf {\bibinfo {volume} {316}},\ \bibinfo
  {pages} {723} (\bibinfo {year} {2007})}\BibitemShut {NoStop}%
\bibitem [{\citenamefont {Buluta1}\ \emph {et~al.}(2011)\citenamefont
  {Buluta1}, \citenamefont {Ashhab1},\ and\ \citenamefont {Nori}}]{Buluta1_11}%
  \BibitemOpen
  \bibfield  {author} {\bibinfo {author} {\bibfnamefont {I.}~\bibnamefont
  {Buluta1}}, \bibinfo {author} {\bibfnamefont {S.}~\bibnamefont {Ashhab1}}, \
  and\ \bibinfo {author} {\bibfnamefont {F.}~\bibnamefont {Nori}},\ }\bibfield
  {title} {\enquote {\bibinfo {title} {Natural and artificial atoms for quantum
  computation},}\ }\href {\doibase 10.1088/0034-4885/74/10/104401} {\bibfield
  {journal} {\bibinfo  {journal} {Rep. Prog. Phys.}\ }\textbf {\bibinfo
  {volume} {74}},\ \bibinfo {pages} {104401} (\bibinfo {year}
  {2011})}\BibitemShut {NoStop}%
\bibitem [{\citenamefont {Xiang}\ \emph {et~al.}(2013)\citenamefont {Xiang},
  \citenamefont {Ashhab}, \citenamefont {You},\ and\ \citenamefont
  {Nori}}]{Xiang_13}%
  \BibitemOpen
  \bibfield  {author} {\bibinfo {author} {\bibfnamefont {Z.~L.}\ \bibnamefont
  {Xiang}}, \bibinfo {author} {\bibfnamefont {S.}~\bibnamefont {Ashhab}},
  \bibinfo {author} {\bibfnamefont {J.~Q.}\ \bibnamefont {You}}, \ and\
  \bibinfo {author} {\bibfnamefont {F.}~\bibnamefont {Nori}},\ }\bibfield
  {title} {\enquote {\bibinfo {title} {Hybrid quantum circuits: Superconducting
  circuits interacting with other quantum systems},}\ }\href {\doibase
  10.1103/RevModPhys.85.623} {\bibfield  {journal} {\bibinfo  {journal} {Rev.
  Mod. Phys.}\ }\textbf {\bibinfo {volume} {85}},\ \bibinfo {pages} {623--653}
  (\bibinfo {year} {2013})}\BibitemShut {NoStop}%
\bibitem [{\citenamefont {Scully}\ and\ \citenamefont
  {Zubairy}(1999)}]{scully_97}%
  \BibitemOpen
  \bibfield  {author} {\bibinfo {author} {\bibfnamefont {M.~O.}\ \bibnamefont
  {Scully}}\ and\ \bibinfo {author} {\bibfnamefont {M.~S.}\ \bibnamefont
  {Zubairy}},\ }\href@noop {} {\enquote {\bibinfo {title} {Quantum optics},}\ }
  (\bibinfo {year} {1999})\BibitemShut {NoStop}%
\bibitem [{\citenamefont {Hood}\ \emph {et~al.}(1998)\citenamefont {Hood},
  \citenamefont {Chapman}, \citenamefont {Lynn},\ and\ \citenamefont
  {Kimble}}]{Hood_98}%
  \BibitemOpen
  \bibfield  {author} {\bibinfo {author} {\bibfnamefont {C.~J.}\ \bibnamefont
  {Hood}}, \bibinfo {author} {\bibfnamefont {M.~S.}\ \bibnamefont {Chapman}},
  \bibinfo {author} {\bibfnamefont {T.~W.}\ \bibnamefont {Lynn}}, \ and\
  \bibinfo {author} {\bibfnamefont {H.~J.}\ \bibnamefont {Kimble}},\ }\bibfield
   {title} {\enquote {\bibinfo {title} {Real-time cavity {QED} with single
  atoms},}\ }\href {\doibase 10.1103/PhysRevLett.80.4157} {\bibfield  {journal}
  {\bibinfo  {journal} {Phys. Rev. Lett.}\ }\textbf {\bibinfo {volume} {80}},\
  \bibinfo {pages} {4157--4160} (\bibinfo {year} {1998})}\BibitemShut {NoStop}%
\bibitem [{\citenamefont {Kimble}(1998)}]{Kimble_98}%
  \BibitemOpen
  \bibfield  {author} {\bibinfo {author} {\bibfnamefont {H.~J.}\ \bibnamefont
  {Kimble}},\ }\bibfield  {title} {\enquote {\bibinfo {title} {Strong
  interactions of single atoms and photons in cavity {QED}},}\ }\href {\doibase
  10.1238/Physica.Topical.076a00127} {\bibfield  {journal} {\bibinfo  {journal}
  {Phys. Scr.}\ }\textbf {\bibinfo {volume} {T76}},\ \bibinfo {pages} {127}
  (\bibinfo {year} {1998})}\BibitemShut {NoStop}%
\bibitem [{\citenamefont {Mabuchi}\ and\ \citenamefont
  {Doherty}(2002)}]{Mabuchi_02}%
  \BibitemOpen
  \bibfield  {author} {\bibinfo {author} {\bibfnamefont {H.}~\bibnamefont
  {Mabuchi}}\ and\ \bibinfo {author} {\bibfnamefont {A.~C.}\ \bibnamefont
  {Doherty}},\ }\bibfield  {title} {\enquote {\bibinfo {title} {Cavity quantum
  electrodynamics: Coherence in context},}\ }\href {\doibase
  10.1126/science.1078446} {\bibfield  {journal} {\bibinfo  {journal}
  {Science}\ }\textbf {\bibinfo {volume} {298}},\ \bibinfo {pages} {1372}
  (\bibinfo {year} {2002})}\BibitemShut {NoStop}%
\bibitem [{\citenamefont {Hennessy}\ \emph {et~al.}(2007)\citenamefont
  {Hennessy}, \citenamefont {Badolato}, \citenamefont {Winger}, \citenamefont
  {Gerace}, \citenamefont {Atat\"{u}re}, \citenamefont {Gulde}, \citenamefont
  {F\"{a}lt}, \citenamefont {Hu},\ and\ \citenamefont
  {Imamo\u{g}lu}}]{Hennessy_07}%
  \BibitemOpen
  \bibfield  {author} {\bibinfo {author} {\bibfnamefont {K.}~\bibnamefont
  {Hennessy}}, \bibinfo {author} {\bibfnamefont {A.}~\bibnamefont {Badolato}},
  \bibinfo {author} {\bibfnamefont {M.}~\bibnamefont {Winger}}, \bibinfo
  {author} {\bibfnamefont {D.}~\bibnamefont {Gerace}}, \bibinfo {author}
  {\bibfnamefont {M.}~\bibnamefont {Atat\"{u}re}}, \bibinfo {author}
  {\bibfnamefont {S.}~\bibnamefont {Gulde}}, \bibinfo {author} {\bibfnamefont
  {S.}~\bibnamefont {F\"{a}lt}}, \bibinfo {author} {\bibfnamefont {E.~L.}\
  \bibnamefont {Hu}}, \ and\ \bibinfo {author} {\bibfnamefont {A.}~\bibnamefont
  {Imamo\u{g}lu}},\ }\bibfield  {title} {\enquote {\bibinfo {title} {Quantum
  nature of a strongly coupled single quantum dot–cavity system},}\ }\href
  {\doibase 10.1038/nature05586} {\bibfield  {journal} {\bibinfo  {journal}
  {Nature}\ }\textbf {\bibinfo {volume} {445}},\ \bibinfo {pages} {896}
  (\bibinfo {year} {2007})}\BibitemShut {NoStop}%
\bibitem [{\citenamefont {You}\ and\ \citenamefont {Nori}(2003)}]{You_03}%
  \BibitemOpen
  \bibfield  {author} {\bibinfo {author} {\bibfnamefont {J.~Q.}\ \bibnamefont
  {You}}\ and\ \bibinfo {author} {\bibfnamefont {F.}~\bibnamefont {Nori}},\
  }\bibfield  {title} {\enquote {\bibinfo {title} {Quantum information
  processing with superconducting qubits in a microwave field},}\ }\href
  {\doibase 10.1103/PhysRevB.68.064509} {\bibfield  {journal} {\bibinfo
  {journal} {Phys. Rev. B}\ }\textbf {\bibinfo {volume} {68}},\ \bibinfo
  {pages} {064509} (\bibinfo {year} {2003})}\BibitemShut {NoStop}%
\bibitem [{\citenamefont {Paik}\ \emph {et~al.}(2011)\citenamefont {Paik},
  \citenamefont {Schuster}, \citenamefont {Bishop}, \citenamefont {Kirchmair},
  \citenamefont {Catelani}, \citenamefont {Sears}, \citenamefont {Johnson},
  \citenamefont {Reagor}, \citenamefont {Frunzio}, \citenamefont {Glazman},
  \citenamefont {Girvin}, \citenamefont {Devoret},\ and\ \citenamefont
  {Schoelkopf}}]{Paik_11}%
  \BibitemOpen
  \bibfield  {author} {\bibinfo {author} {\bibfnamefont {H.}~\bibnamefont
  {Paik}}, \bibinfo {author} {\bibfnamefont {D.~I.}\ \bibnamefont {Schuster}},
  \bibinfo {author} {\bibfnamefont {Lev~S.}\ \bibnamefont {Bishop}}, \bibinfo
  {author} {\bibfnamefont {G.}~\bibnamefont {Kirchmair}}, \bibinfo {author}
  {\bibfnamefont {G.}~\bibnamefont {Catelani}}, \bibinfo {author}
  {\bibfnamefont {A.~P.}\ \bibnamefont {Sears}}, \bibinfo {author}
  {\bibfnamefont {B.~R.}\ \bibnamefont {Johnson}}, \bibinfo {author}
  {\bibfnamefont {M.~J.}\ \bibnamefont {Reagor}}, \bibinfo {author}
  {\bibfnamefont {L.}~\bibnamefont {Frunzio}}, \bibinfo {author} {\bibfnamefont
  {L.~I.}\ \bibnamefont {Glazman}}, \bibinfo {author} {\bibfnamefont {S.~M.}\
  \bibnamefont {Girvin}}, \bibinfo {author} {\bibfnamefont {M.~H.}\
  \bibnamefont {Devoret}}, \ and\ \bibinfo {author} {\bibfnamefont {R.~J.}\
  \bibnamefont {Schoelkopf}},\ }\bibfield  {title} {\enquote {\bibinfo {title}
  {Observation of high coherence in josephson junction qubits measured in a
  three-dimensional circuit {QED} architecture},}\ }\href {\doibase
  10.1103/PhysRevLett.107.240501} {\bibfield  {journal} {\bibinfo  {journal}
  {Phys. Rev. Lett.}\ }\textbf {\bibinfo {volume} {107}},\ \bibinfo {pages}
  {240501} (\bibinfo {year} {2011})}\BibitemShut {NoStop}%
\bibitem [{\citenamefont {Stern}\ \emph {et~al.}(2014)\citenamefont {Stern},
  \citenamefont {Catelani}, \citenamefont {Kubo}, \citenamefont {Grezes},
  \citenamefont {Bienfait}, \citenamefont {Vion}, \citenamefont {Esteve},\ and\
  \citenamefont {Bertet}}]{Stern_14}%
  \BibitemOpen
  \bibfield  {author} {\bibinfo {author} {\bibfnamefont {M.}~\bibnamefont
  {Stern}}, \bibinfo {author} {\bibfnamefont {G.}~\bibnamefont {Catelani}},
  \bibinfo {author} {\bibfnamefont {Y.}~\bibnamefont {Kubo}}, \bibinfo {author}
  {\bibfnamefont {C.}~\bibnamefont {Grezes}}, \bibinfo {author} {\bibfnamefont
  {A.}~\bibnamefont {Bienfait}}, \bibinfo {author} {\bibfnamefont
  {D.}~\bibnamefont {Vion}}, \bibinfo {author} {\bibfnamefont {D.}~\bibnamefont
  {Esteve}}, \ and\ \bibinfo {author} {\bibfnamefont {P.}~\bibnamefont
  {Bertet}},\ }\bibfield  {title} {\enquote {\bibinfo {title} {Flux qubits with
  long coherence times for hybrid quantum circuits},}\ }\href {\doibase
  10.1103/PhysRevLett.113.123601} {\bibfield  {journal} {\bibinfo  {journal}
  {Phys. Rev. Lett.}\ }\textbf {\bibinfo {volume} {113}},\ \bibinfo {pages}
  {123601} (\bibinfo {year} {2014})}\BibitemShut {NoStop}%
\bibitem [{\citenamefont {Niemczyk}\ \emph {et~al.}(2010)\citenamefont
  {Niemczyk}, \citenamefont {Deppe}, \citenamefont {Huebl}, \citenamefont
  {Menzel}, \citenamefont {Hocke}, \citenamefont {Schwarz}, \citenamefont
  {Garcia-Ripoll}, \citenamefont {Zueco}, \citenamefont {H\"{u}mmer},
  \citenamefont {Solano}, \citenamefont {Marx},\ and\ \citenamefont
  {Gross}}]{Niemczyk_10}%
  \BibitemOpen
  \bibfield  {author} {\bibinfo {author} {\bibfnamefont {T.}~\bibnamefont
  {Niemczyk}}, \bibinfo {author} {\bibfnamefont {F.}~\bibnamefont {Deppe}},
  \bibinfo {author} {\bibfnamefont {H.}~\bibnamefont {Huebl}}, \bibinfo
  {author} {\bibfnamefont {E.~P.}\ \bibnamefont {Menzel}}, \bibinfo {author}
  {\bibfnamefont {F.}~\bibnamefont {Hocke}}, \bibinfo {author} {\bibfnamefont
  {M.~J.}\ \bibnamefont {Schwarz}}, \bibinfo {author} {\bibfnamefont {J.~J.}\
  \bibnamefont {Garcia-Ripoll}}, \bibinfo {author} {\bibfnamefont
  {D.}~\bibnamefont {Zueco}}, \bibinfo {author} {\bibfnamefont
  {T.}~\bibnamefont {H\"{u}mmer}}, \bibinfo {author} {\bibfnamefont
  {E.}~\bibnamefont {Solano}}, \bibinfo {author} {\bibfnamefont
  {A.}~\bibnamefont {Marx}}, \ and\ \bibinfo {author} {\bibfnamefont
  {R.}~\bibnamefont {Gross}},\ }\bibfield  {title} {\enquote {\bibinfo {title}
  {Circuit quantum electrodynamics in the ultrastrong-coupling regime},}\
  }\href {\doibase 10.1038/nphys1730} {\bibfield  {journal} {\bibinfo
  {journal} {Nat. Phys.}\ }\textbf {\bibinfo {volume} {6}},\ \bibinfo {pages}
  {772} (\bibinfo {year} {2010})}\BibitemShut {NoStop}%
\bibitem [{\citenamefont {Forn-D\'{\i}az}\ \emph {et~al.}(2010)\citenamefont
  {Forn-D\'{\i}az}, \citenamefont {Lisenfeld}, \citenamefont {Marcos},
  \citenamefont {Garc\'{\i}a-Ripoll}, \citenamefont {Solano}, \citenamefont
  {Harmans},\ and\ \citenamefont {Mooij}}]{Forn_10}%
  \BibitemOpen
  \bibfield  {author} {\bibinfo {author} {\bibfnamefont {P.}~\bibnamefont
  {Forn-D\'{\i}az}}, \bibinfo {author} {\bibfnamefont {J.}~\bibnamefont
  {Lisenfeld}}, \bibinfo {author} {\bibfnamefont {D.}~\bibnamefont {Marcos}},
  \bibinfo {author} {\bibfnamefont {J.~J.}\ \bibnamefont {Garc\'{\i}a-Ripoll}},
  \bibinfo {author} {\bibfnamefont {E.}~\bibnamefont {Solano}}, \bibinfo
  {author} {\bibfnamefont {C.~J. P.~M.}\ \bibnamefont {Harmans}}, \ and\
  \bibinfo {author} {\bibfnamefont {J.~E.}\ \bibnamefont {Mooij}},\ }\bibfield
  {title} {\enquote {\bibinfo {title} {Observation of the bloch-siegert shift
  in a qubit-oscillator system in the ultrastrong coupling regime},}\ }\href
  {\doibase 10.1103/PhysRevLett.105.237001} {\bibfield  {journal} {\bibinfo
  {journal} {Phys. Rev. Lett.}\ }\textbf {\bibinfo {volume} {105}},\ \bibinfo
  {pages} {237001} (\bibinfo {year} {2010})}\BibitemShut {NoStop}%
\bibitem [{\citenamefont {Yoshihara}\ \emph {et~al.}(2017)\citenamefont
  {Yoshihara}, \citenamefont {Fuse}, \citenamefont {Ashhab}, \citenamefont
  {Kakuyanagi}, \citenamefont {Saito},\ and\ \citenamefont
  {Semba}}]{Yoshihara_17}%
  \BibitemOpen
  \bibfield  {author} {\bibinfo {author} {\bibfnamefont {F.}~\bibnamefont
  {Yoshihara}}, \bibinfo {author} {\bibfnamefont {T.}~\bibnamefont {Fuse}},
  \bibinfo {author} {\bibfnamefont {S.}~\bibnamefont {Ashhab}}, \bibinfo
  {author} {\bibfnamefont {K.}~\bibnamefont {Kakuyanagi}}, \bibinfo {author}
  {\bibfnamefont {S.}~\bibnamefont {Saito}}, \ and\ \bibinfo {author}
  {\bibfnamefont {K.}~\bibnamefont {Semba}},\ }\bibfield  {title} {\enquote
  {\bibinfo {title} {Superconducting qubit–oscillator circuit beyond the
  ultrastrong-coupling regime},}\ }\href {\doibase 10.1038/nphys3906}
  {\bibfield  {journal} {\bibinfo  {journal} {Nat. Phys.}\ }\textbf {\bibinfo
  {volume} {13}},\ \bibinfo {pages} {44} (\bibinfo {year} {2017})}\BibitemShut
  {NoStop}%
\bibitem [{\citenamefont {Braum\"{u}ller}\ \emph {et~al.}(2017)\citenamefont
  {Braum\"{u}ller}, \citenamefont {Marthaler}, \citenamefont {Schneider},
  \citenamefont {Stehli}, \citenamefont {Rotzinger}, \citenamefont {Weides},\
  and\ \citenamefont {Ustinov}}]{Braumueller_17}%
  \BibitemOpen
  \bibfield  {author} {\bibinfo {author} {\bibfnamefont {J.}~\bibnamefont
  {Braum\"{u}ller}}, \bibinfo {author} {\bibfnamefont {M.}~\bibnamefont
  {Marthaler}}, \bibinfo {author} {\bibfnamefont {A.}~\bibnamefont
  {Schneider}}, \bibinfo {author} {\bibfnamefont {A.}~\bibnamefont {Stehli}},
  \bibinfo {author} {\bibfnamefont {H.}~\bibnamefont {Rotzinger}}, \bibinfo
  {author} {\bibfnamefont {M.}~\bibnamefont {Weides}}, \ and\ \bibinfo {author}
  {\bibfnamefont {A.~V.}\ \bibnamefont {Ustinov}},\ }\bibfield  {title}
  {\enquote {\bibinfo {title} {Analog quantum simulation of the {R}abi model in
  the ultra-strong coupling regime},}\ }\href {\doibase
  10.1038/s41467-017-00894-w} {\bibfield  {journal} {\bibinfo  {journal} {Nat.
  Commun.}\ }\textbf {\bibinfo {volume} {8}},\ \bibinfo {pages} {779} (\bibinfo
  {year} {2017})}\BibitemShut {NoStop}%
\bibitem [{\citenamefont {Forn-D\'{\i}az}\ \emph {et~al.}(2017)\citenamefont
  {Forn-D\'{\i}az}, \citenamefont {Garc\'{\i}a-Ripoll}, \citenamefont
  {Peropadre}, \citenamefont {Orgiazzi}, \citenamefont {Belyansky},
  \citenamefont {Wilson},\ and\ \citenamefont {Lupascu}}]{Forn_17}%
  \BibitemOpen
  \bibfield  {author} {\bibinfo {author} {\bibfnamefont {P.}~\bibnamefont
  {Forn-D\'{\i}az}}, \bibinfo {author} {\bibfnamefont {J.~J.}\ \bibnamefont
  {Garc\'{\i}a-Ripoll}}, \bibinfo {author} {\bibfnamefont {B.}~\bibnamefont
  {Peropadre}}, \bibinfo {author} {\bibfnamefont {M.~A.}\ \bibnamefont
  {Orgiazzi}, \bibfnamefont {J.~L.~andYurtalan}}, \bibinfo {author}
  {\bibfnamefont {R.}~\bibnamefont {Belyansky}}, \bibinfo {author}
  {\bibfnamefont {C.~M.}\ \bibnamefont {Wilson}}, \ and\ \bibinfo {author}
  {\bibfnamefont {A.}~\bibnamefont {Lupascu}},\ }\bibfield  {title} {\enquote
  {\bibinfo {title} {Ultrastrong coupling of a single artificial atom to an
  electromagnetic continuum in the nonperturbative regime},}\ }\href {\doibase
  10.1038/nphys3905} {\bibfield  {journal} {\bibinfo  {journal} {Nat. Phys.}\
  }\textbf {\bibinfo {volume} {13}},\ \bibinfo {pages} {39} (\bibinfo {year}
  {2017})}\BibitemShut {NoStop}%
\bibitem [{\citenamefont {Nataf}\ and\ \citenamefont {Ciuti}(2010)}]{Nataf_10}%
  \BibitemOpen
  \bibfield  {author} {\bibinfo {author} {\bibfnamefont {P.}~\bibnamefont
  {Nataf}}\ and\ \bibinfo {author} {\bibfnamefont {C.}~\bibnamefont {Ciuti}},\
  }\bibfield  {title} {\enquote {\bibinfo {title} {Vacuum degeneracy of a
  circuit {QED} system in the ultrastrong coupling regime},}\ }\href {\doibase
  10.1103/PhysRevLett.104.023601} {\bibfield  {journal} {\bibinfo  {journal}
  {Phys. Rev. Lett.}\ }\textbf {\bibinfo {volume} {104}},\ \bibinfo {pages}
  {023601} (\bibinfo {year} {2010})}\BibitemShut {NoStop}%
\bibitem [{\citenamefont {Peropadre}\ \emph {et~al.}(2010)\citenamefont
  {Peropadre}, \citenamefont {Forn-D\'{\i}az}, \citenamefont {Solano},\ and\
  \citenamefont {Garc\'{\i}a-Ripoll}}]{Peropadre_10}%
  \BibitemOpen
  \bibfield  {author} {\bibinfo {author} {\bibfnamefont {B.}~\bibnamefont
  {Peropadre}}, \bibinfo {author} {\bibfnamefont {P.}~\bibnamefont
  {Forn-D\'{\i}az}}, \bibinfo {author} {\bibfnamefont {E.}~\bibnamefont
  {Solano}}, \ and\ \bibinfo {author} {\bibfnamefont {J.~J.}\ \bibnamefont
  {Garc\'{\i}a-Ripoll}},\ }\bibfield  {title} {\enquote {\bibinfo {title}
  {Switchable ultrastrong coupling in circuit {QED}},}\ }\href {\doibase
  10.1103/PhysRevLett.105.023601} {\bibfield  {journal} {\bibinfo  {journal}
  {Phys. Rev. Lett.}\ }\textbf {\bibinfo {volume} {105}},\ \bibinfo {pages}
  {023601} (\bibinfo {year} {2010})}\BibitemShut {NoStop}%
\bibitem [{\citenamefont {Ridolfo}\ \emph {et~al.}(2012)\citenamefont
  {Ridolfo}, \citenamefont {Leib}, \citenamefont {Savasta},\ and\ \citenamefont
  {Hartmann}}]{Ridolfo_12}%
  \BibitemOpen
  \bibfield  {author} {\bibinfo {author} {\bibfnamefont {A.}~\bibnamefont
  {Ridolfo}}, \bibinfo {author} {\bibfnamefont {M.}~\bibnamefont {Leib}},
  \bibinfo {author} {\bibfnamefont {S.}~\bibnamefont {Savasta}}, \ and\
  \bibinfo {author} {\bibfnamefont {M.~J.}\ \bibnamefont {Hartmann}},\
  }\bibfield  {title} {\enquote {\bibinfo {title} {Photon blockade in the
  ultrastrong coupling regime},}\ }\href {\doibase
  10.1103/PhysRevLett.109.193602} {\bibfield  {journal} {\bibinfo  {journal}
  {Phys. Rev. Lett.}\ }\textbf {\bibinfo {volume} {109}},\ \bibinfo {pages}
  {193602} (\bibinfo {year} {2012})}\BibitemShut {NoStop}%
\bibitem [{\citenamefont {Ridolfo}\ \emph {et~al.}(2013)\citenamefont
  {Ridolfo}, \citenamefont {Savasta},\ and\ \citenamefont
  {Hartmann}}]{Ridolfo_13}%
  \BibitemOpen
  \bibfield  {author} {\bibinfo {author} {\bibfnamefont {A.}~\bibnamefont
  {Ridolfo}}, \bibinfo {author} {\bibfnamefont {S.}~\bibnamefont {Savasta}}, \
  and\ \bibinfo {author} {\bibfnamefont {M.~J.}\ \bibnamefont {Hartmann}},\
  }\bibfield  {title} {\enquote {\bibinfo {title} {Nonclassical radiation from
  thermal cavities in the ultrastrong coupling regime},}\ }\href {\doibase
  10.1103/PhysRevLett.110.163601} {\bibfield  {journal} {\bibinfo  {journal}
  {Phys. Rev. Lett.}\ }\textbf {\bibinfo {volume} {110}},\ \bibinfo {pages}
  {163601} (\bibinfo {year} {2013})}\BibitemShut {NoStop}%
\bibitem [{\citenamefont {Stassi}\ \emph {et~al.}(2013)\citenamefont {Stassi},
  \citenamefont {Ridolfo}, \citenamefont {Di~Stefano}, \citenamefont
  {Hartmann},\ and\ \citenamefont {Savasta}}]{Stassi_13}%
  \BibitemOpen
  \bibfield  {author} {\bibinfo {author} {\bibfnamefont {R.}~\bibnamefont
  {Stassi}}, \bibinfo {author} {\bibfnamefont {A.}~\bibnamefont {Ridolfo}},
  \bibinfo {author} {\bibfnamefont {O.}~\bibnamefont {Di~Stefano}}, \bibinfo
  {author} {\bibfnamefont {M.~J.}\ \bibnamefont {Hartmann}}, \ and\ \bibinfo
  {author} {\bibfnamefont {S.}~\bibnamefont {Savasta}},\ }\bibfield  {title}
  {\enquote {\bibinfo {title} {Spontaneous conversion from virtual to real
  photons in the ultrastrong-coupling regime},}\ }\href {\doibase
  10.1103/PhysRevLett.110.243601} {\bibfield  {journal} {\bibinfo  {journal}
  {Phys. Rev. Lett.}\ }\textbf {\bibinfo {volume} {110}},\ \bibinfo {pages}
  {243601} (\bibinfo {year} {2013})}\BibitemShut {NoStop}%
\bibitem [{\citenamefont {Garziano}\ \emph {et~al.}(2014)\citenamefont
  {Garziano}, \citenamefont {Stassi}, \citenamefont {Ridolfo}, \citenamefont
  {Di~Stefano},\ and\ \citenamefont {Savasta}}]{Garziano_14}%
  \BibitemOpen
  \bibfield  {author} {\bibinfo {author} {\bibfnamefont {L.}~\bibnamefont
  {Garziano}}, \bibinfo {author} {\bibfnamefont {R.}~\bibnamefont {Stassi}},
  \bibinfo {author} {\bibfnamefont {A.}~\bibnamefont {Ridolfo}}, \bibinfo
  {author} {\bibfnamefont {O.}~\bibnamefont {Di~Stefano}}, \ and\ \bibinfo
  {author} {\bibfnamefont {S.}~\bibnamefont {Savasta}},\ }\bibfield  {title}
  {\enquote {\bibinfo {title} {Vacuum-induced symmetry breaking in a
  superconducting quantum circuit},}\ }\href {\doibase
  10.1103/PhysRevA.90.043817} {\bibfield  {journal} {\bibinfo  {journal} {Phys.
  Rev. A}\ }\textbf {\bibinfo {volume} {90}},\ \bibinfo {pages} {043817}
  (\bibinfo {year} {2014})}\BibitemShut {NoStop}%
\bibitem [{\citenamefont {Garziano}\ \emph {et~al.}(2015)\citenamefont
  {Garziano}, \citenamefont {Stassi}, \citenamefont {Macr\`{\i}}, \citenamefont
  {Kockum}, \citenamefont {Savasta},\ and\ \citenamefont {Nori}}]{Garziano_15}%
  \BibitemOpen
  \bibfield  {author} {\bibinfo {author} {\bibfnamefont {L.}~\bibnamefont
  {Garziano}}, \bibinfo {author} {\bibfnamefont {R.}~\bibnamefont {Stassi}},
  \bibinfo {author} {\bibfnamefont {V.}~\bibnamefont {Macr\`{\i}}}, \bibinfo
  {author} {\bibfnamefont {A.~F.}\ \bibnamefont {Kockum}}, \bibinfo {author}
  {\bibfnamefont {S.}~\bibnamefont {Savasta}}, \ and\ \bibinfo {author}
  {\bibfnamefont {F.}~\bibnamefont {Nori}},\ }\bibfield  {title} {\enquote
  {\bibinfo {title} {Multiphoton quantum {R}abi oscillations in ultrastrong
  cavity {QED}},}\ }\href {\doibase 10.1103/PhysRevA.92.063830} {\bibfield
  {journal} {\bibinfo  {journal} {Phys. Rev. A}\ }\textbf {\bibinfo {volume}
  {92}},\ \bibinfo {pages} {063830} (\bibinfo {year} {2015})}\BibitemShut
  {NoStop}%
\bibitem [{\citenamefont {Ma}\ and\ \citenamefont {Law}(2015)}]{Ma_15}%
  \BibitemOpen
  \bibfield  {author} {\bibinfo {author} {\bibfnamefont {Ken~K.~W.}\
  \bibnamefont {Ma}}\ and\ \bibinfo {author} {\bibfnamefont {C.~K.}\
  \bibnamefont {Law}},\ }\bibfield  {title} {\enquote {\bibinfo {title}
  {Three-photon resonance and adiabatic passage in the large-detuning {R}abi
  model},}\ }\href {\doibase 10.1103/PhysRevA.92.023842} {\bibfield  {journal}
  {\bibinfo  {journal} {Phys. Rev. A}\ }\textbf {\bibinfo {volume} {92}},\
  \bibinfo {pages} {023842} (\bibinfo {year} {2015})}\BibitemShut {NoStop}%
\bibitem [{\citenamefont {Garziano}\ \emph {et~al.}(2016)\citenamefont
  {Garziano}, \citenamefont {Macr\`{\i}}, \citenamefont {Stassi}, \citenamefont
  {Di~Stefano}, \citenamefont {Nori},\ and\ \citenamefont
  {Savasta}}]{Garziano_16}%
  \BibitemOpen
  \bibfield  {author} {\bibinfo {author} {\bibfnamefont {L.}~\bibnamefont
  {Garziano}}, \bibinfo {author} {\bibfnamefont {V.}~\bibnamefont
  {Macr\`{\i}}}, \bibinfo {author} {\bibfnamefont {R.}~\bibnamefont {Stassi}},
  \bibinfo {author} {\bibfnamefont {O.}~\bibnamefont {Di~Stefano}}, \bibinfo
  {author} {\bibfnamefont {F.}~\bibnamefont {Nori}}, \ and\ \bibinfo {author}
  {\bibfnamefont {S.}~\bibnamefont {Savasta}},\ }\bibfield  {title} {\enquote
  {\bibinfo {title} {One photon can simultaneously excite two or more atoms},}\
  }\href {\doibase 10.1103/PhysRevLett.117.043601} {\bibfield  {journal}
  {\bibinfo  {journal} {Phys. Rev. Lett.}\ }\textbf {\bibinfo {volume} {117}},\
  \bibinfo {pages} {043601} (\bibinfo {year} {2016})}\BibitemShut {NoStop}%
\bibitem [{\citenamefont {Kockum}\ \emph {et~al.}(2017)\citenamefont {Kockum},
  \citenamefont {Macr\`{\i}}, \citenamefont {Garziano}, \citenamefont
  {Savasta},\ and\ \citenamefont {Nori}}]{Anton_17}%
  \BibitemOpen
  \bibfield  {author} {\bibinfo {author} {\bibfnamefont {A.~F.}\ \bibnamefont
  {Kockum}}, \bibinfo {author} {\bibfnamefont {V.}~\bibnamefont {Macr\`{\i}}},
  \bibinfo {author} {\bibfnamefont {L.}~\bibnamefont {Garziano}}, \bibinfo
  {author} {\bibfnamefont {S.}~\bibnamefont {Savasta}}, \ and\ \bibinfo
  {author} {\bibfnamefont {F.}~\bibnamefont {Nori}},\ }\bibfield  {title}
  {\enquote {\bibinfo {title} {Frequency conversion in ultrastrong cavity
  {QED}},}\ }\href {\doibase 10.1038/s41598-017-04225-3} {\bibfield  {journal}
  {\bibinfo  {journal} {Sci. Rep.}\ }\textbf {\bibinfo {volume} {7}},\ \bibinfo
  {pages} {5313} (\bibinfo {year} {2017})}\BibitemShut {NoStop}%
\bibitem [{\citenamefont {Houck}\ \emph {et~al.}(2007)\citenamefont {Houck},
  \citenamefont {Schuster}, \citenamefont {Gambetta}, \citenamefont {Schreier},
  \citenamefont {Johnson}, \citenamefont {Chow}, \citenamefont {Frunzio},
  \citenamefont {Majer}, \citenamefont {Devoret}, \citenamefont {Girvin},\ and\
  \citenamefont {Schoelkopf}}]{Houck_07}%
  \BibitemOpen
  \bibfield  {author} {\bibinfo {author} {\bibfnamefont {A.~A.}\ \bibnamefont
  {Houck}}, \bibinfo {author} {\bibfnamefont {D.~I.}\ \bibnamefont {Schuster}},
  \bibinfo {author} {\bibfnamefont {J.~M.}\ \bibnamefont {Gambetta}}, \bibinfo
  {author} {\bibfnamefont {J.~A.}\ \bibnamefont {Schreier}}, \bibinfo {author}
  {\bibfnamefont {B.~R.}\ \bibnamefont {Johnson}}, \bibinfo {author}
  {\bibfnamefont {J.~M.}\ \bibnamefont {Chow}}, \bibinfo {author}
  {\bibfnamefont {L.}~\bibnamefont {Frunzio}}, \bibinfo {author} {\bibfnamefont
  {J.}~\bibnamefont {Majer}}, \bibinfo {author} {\bibfnamefont {M.~H.}\
  \bibnamefont {Devoret}}, \bibinfo {author} {\bibfnamefont {S.~M.}\
  \bibnamefont {Girvin}}, \ and\ \bibinfo {author} {\bibfnamefont {R.~J.}\
  \bibnamefont {Schoelkopf}},\ }\bibfield  {title} {\enquote {\bibinfo {title}
  {Generating single microwave photons in a circuit},}\ }\href {\doibase
  10.1038/nature06126} {\bibfield  {journal} {\bibinfo  {journal} {Nature}\
  }\textbf {\bibinfo {volume} {449}},\ \bibinfo {pages} {328} (\bibinfo {year}
  {2007})}\BibitemShut {NoStop}%
\bibitem [{\citenamefont {Romero}\ \emph {et~al.}(2009)\citenamefont {Romero},
  \citenamefont {Garc\'{\i}a-Ripoll},\ and\ \citenamefont
  {Solano}}]{Romero_09}%
  \BibitemOpen
  \bibfield  {author} {\bibinfo {author} {\bibfnamefont {G.}~\bibnamefont
  {Romero}}, \bibinfo {author} {\bibfnamefont {J.~J.}\ \bibnamefont
  {Garc\'{\i}a-Ripoll}}, \ and\ \bibinfo {author} {\bibfnamefont
  {E.}~\bibnamefont {Solano}},\ }\bibfield  {title} {\enquote {\bibinfo {title}
  {Microwave photon detector in circuit {QED}},}\ }\href {\doibase
  10.1103/PhysRevLett.102.173602} {\bibfield  {journal} {\bibinfo  {journal}
  {Phys. Rev. Lett.}\ }\textbf {\bibinfo {volume} {102}},\ \bibinfo {pages}
  {173602} (\bibinfo {year} {2009})}\BibitemShut {NoStop}%
\bibitem [{\citenamefont {Peropadre}\ \emph {et~al.}(2011)\citenamefont
  {Peropadre}, \citenamefont {Romero}, \citenamefont {Johansson}, \citenamefont
  {Wilson}, \citenamefont {Solano},\ and\ \citenamefont
  {Garc\'{\i}a-Ripoll}}]{Peropadre_11}%
  \BibitemOpen
  \bibfield  {author} {\bibinfo {author} {\bibfnamefont {B.}~\bibnamefont
  {Peropadre}}, \bibinfo {author} {\bibfnamefont {G.}~\bibnamefont {Romero}},
  \bibinfo {author} {\bibfnamefont {G.}~\bibnamefont {Johansson}}, \bibinfo
  {author} {\bibfnamefont {C.~M.}\ \bibnamefont {Wilson}}, \bibinfo {author}
  {\bibfnamefont {E.}~\bibnamefont {Solano}}, \ and\ \bibinfo {author}
  {\bibfnamefont {J.~J.}\ \bibnamefont {Garc\'{\i}a-Ripoll}},\ }\bibfield
  {title} {\enquote {\bibinfo {title} {Approaching perfect microwave
  photodetection in circuit {QED}},}\ }\href {\doibase
  10.1103/PhysRevA.84.063834} {\bibfield  {journal} {\bibinfo  {journal} {Phys.
  Rev. A}\ }\textbf {\bibinfo {volume} {84}},\ \bibinfo {pages} {063834}
  (\bibinfo {year} {2011})}\BibitemShut {NoStop}%
\bibitem [{\citenamefont {Shen}\ and\ \citenamefont {Fan}(2005)}]{Shen_05}%
  \BibitemOpen
  \bibfield  {author} {\bibinfo {author} {\bibfnamefont {J.~T.}\ \bibnamefont
  {Shen}}\ and\ \bibinfo {author} {\bibfnamefont {S.}~\bibnamefont {Fan}},\
  }\bibfield  {title} {\enquote {\bibinfo {title} {Coherent photon transport
  from spontaneous emission in one-dimensional waveguides},}\ }\href {\doibase
  10.1364/OL.30.002001} {\bibfield  {journal} {\bibinfo  {journal} {Opt.
  Lett.}\ }\textbf {\bibinfo {volume} {30}},\ \bibinfo {pages} {2001--2003}
  (\bibinfo {year} {2005})}\BibitemShut {NoStop}%
\bibitem [{\citenamefont {Zhou}\ \emph {et~al.}(2008)\citenamefont {Zhou},
  \citenamefont {Gong}, \citenamefont {Liu}, \citenamefont {Sun},\ and\
  \citenamefont {Nori}}]{Zhou_08}%
  \BibitemOpen
  \bibfield  {author} {\bibinfo {author} {\bibfnamefont {L.}~\bibnamefont
  {Zhou}}, \bibinfo {author} {\bibfnamefont {Z.~R.}\ \bibnamefont {Gong}},
  \bibinfo {author} {\bibfnamefont {Y.~X.}\ \bibnamefont {Liu}}, \bibinfo
  {author} {\bibfnamefont {C.~P.}\ \bibnamefont {Sun}}, \ and\ \bibinfo
  {author} {\bibfnamefont {F.}~\bibnamefont {Nori}},\ }\bibfield  {title}
  {\enquote {\bibinfo {title} {Controllable scattering of a single photon
  inside a one-dimensional resonator waveguide},}\ }\href {\doibase
  10.1103/PhysRevLett.101.100501} {\bibfield  {journal} {\bibinfo  {journal}
  {Phys. Rev. Lett.}\ }\textbf {\bibinfo {volume} {101}},\ \bibinfo {pages}
  {100501} (\bibinfo {year} {2008})}\BibitemShut {NoStop}%
\bibitem [{\citenamefont {Shen}\ and\ \citenamefont {Fan}(2009)}]{JTShen_09}%
  \BibitemOpen
  \bibfield  {author} {\bibinfo {author} {\bibfnamefont {J.~T.}\ \bibnamefont
  {Shen}}\ and\ \bibinfo {author} {\bibfnamefont {S.~H.}\ \bibnamefont {Fan}},\
  }\bibfield  {title} {\enquote {\bibinfo {title} {Theory of single-photon
  transport in a single-mode waveguide. {I}. {C}oupling to a cavity containing
  a two-level atom},}\ }\href {\doibase 10.1103/PhysRevA.79.023837} {\bibfield
  {journal} {\bibinfo  {journal} {Phys. Rev. A}\ }\textbf {\bibinfo {volume}
  {79}},\ \bibinfo {pages} {023837} (\bibinfo {year} {2009})}\BibitemShut
  {NoStop}%
\bibitem [{\citenamefont {Witthaut}\ and\ \citenamefont
  {Sorensen}(2010)}]{Witthaut_10}%
  \BibitemOpen
  \bibfield  {author} {\bibinfo {author} {\bibfnamefont {D.}~\bibnamefont
  {Witthaut}}\ and\ \bibinfo {author} {\bibfnamefont {A.~S.}\ \bibnamefont
  {Sorensen}},\ }\bibfield  {title} {\enquote {\bibinfo {title} {Photon
  scattering by a three-level emitter in a one-dimensional waveguide},}\ }\href
  {\doibase 10.1088/1367-2630/12/4/043052} {\bibfield  {journal} {\bibinfo
  {journal} {New J. Phys.}\ }\textbf {\bibinfo {volume} {12}},\ \bibinfo
  {pages} {043052} (\bibinfo {year} {2010})}\BibitemShut {NoStop}%
\bibitem [{\citenamefont {Chen}\ \emph {et~al.}(2014)\citenamefont {Chen},
  \citenamefont {Liu},\ and\ \citenamefont {Chen}}]{GYChen_14}%
  \BibitemOpen
  \bibfield  {author} {\bibinfo {author} {\bibfnamefont {G.~Y.}\ \bibnamefont
  {Chen}}, \bibinfo {author} {\bibfnamefont {M.~H.}\ \bibnamefont {Liu}}, \
  and\ \bibinfo {author} {\bibfnamefont {Y.~N.}\ \bibnamefont {Chen}},\
  }\bibfield  {title} {\enquote {\bibinfo {title} {Scattering of microwave
  photons in superconducting transmission-line resonators coupled to charge
  qubits},}\ }\href {\doibase 10.1103/PhysRevA.89.053802} {\bibfield  {journal}
  {\bibinfo  {journal} {Phys. Rev. A}\ }\textbf {\bibinfo {volume} {89}},\
  \bibinfo {pages} {053802} (\bibinfo {year} {2014})}\BibitemShut {NoStop}%
\bibitem [{\citenamefont {Roy}\ \emph {et~al.}(2017)\citenamefont {Roy},
  \citenamefont {Wilson},\ and\ \citenamefont {Firstenberg}}]{Roy_17}%
  \BibitemOpen
  \bibfield  {author} {\bibinfo {author} {\bibfnamefont {D.}~\bibnamefont
  {Roy}}, \bibinfo {author} {\bibfnamefont {C.~M.}\ \bibnamefont {Wilson}}, \
  and\ \bibinfo {author} {\bibfnamefont {O.}~\bibnamefont {Firstenberg}},\
  }\bibfield  {title} {\enquote {\bibinfo {title} {Colloquium: Strongly
  interacting photons in one-dimensional continuum},}\ }\href {\doibase
  10.1103/RevModPhys.89.021001} {\bibfield  {journal} {\bibinfo  {journal}
  {Rev. Mod. Phys.}\ }\textbf {\bibinfo {volume} {89}},\ \bibinfo {pages}
  {021001} (\bibinfo {year} {2017})}\BibitemShut {NoStop}%
\bibitem [{\citenamefont {Chang}\ \emph {et~al.}(2007)\citenamefont {Chang},
  \citenamefont {Sorensen}, \citenamefont {Demler},\ and\ \citenamefont
  {Lukin}}]{Chang_07}%
  \BibitemOpen
  \bibfield  {author} {\bibinfo {author} {\bibfnamefont {D.~E.}\ \bibnamefont
  {Chang}}, \bibinfo {author} {\bibfnamefont {A.~S.}\ \bibnamefont {Sorensen}},
  \bibinfo {author} {\bibfnamefont {E.~A.}\ \bibnamefont {Demler}}, \ and\
  \bibinfo {author} {\bibfnamefont {M.~D.}\ \bibnamefont {Lukin}},\ }\bibfield
  {title} {\enquote {\bibinfo {title} {A single-photon transistor using
  nanoscale surface plasmons},}\ }\href {\doibase 10.1038/nphys708} {\bibfield
  {journal} {\bibinfo  {journal} {Nat. Phys.}\ }\textbf {\bibinfo {volume}
  {3}},\ \bibinfo {pages} {807} (\bibinfo {year} {2007})}\BibitemShut {NoStop}%
\bibitem [{\citenamefont {Chen}\ \emph
  {et~al.}(2011{\natexlab{a}})\citenamefont {Chen}, \citenamefont {Lambert},
  \citenamefont {Chou}, \citenamefont {Chen},\ and\ \citenamefont
  {Nori}}]{GYChen_11}%
  \BibitemOpen
  \bibfield  {author} {\bibinfo {author} {\bibfnamefont {G.~Y.}\ \bibnamefont
  {Chen}}, \bibinfo {author} {\bibfnamefont {N.}~\bibnamefont {Lambert}},
  \bibinfo {author} {\bibfnamefont {C.~H.}\ \bibnamefont {Chou}}, \bibinfo
  {author} {\bibfnamefont {Y.~N.}\ \bibnamefont {Chen}}, \ and\ \bibinfo
  {author} {\bibfnamefont {F.}~\bibnamefont {Nori}},\ }\bibfield  {title}
  {\enquote {\bibinfo {title} {Surface plasmons in a metal nanowire coupled to
  colloidal quantum dots: Scattering properties and quantum entanglement},}\
  }\href {\doibase 10.1103/PhysRevB.84.045310} {\bibfield  {journal} {\bibinfo
  {journal} {Phys. Rev. B}\ }\textbf {\bibinfo {volume} {84}},\ \bibinfo
  {pages} {045310} (\bibinfo {year} {2011}{\natexlab{a}})}\BibitemShut
  {NoStop}%
\bibitem [{\citenamefont {Chen}\ \emph
  {et~al.}(2011{\natexlab{b}})\citenamefont {Chen}, \citenamefont {Chen},\ and\
  \citenamefont {Chen}}]{WChen_11}%
  \BibitemOpen
  \bibfield  {author} {\bibinfo {author} {\bibfnamefont {W.}~\bibnamefont
  {Chen}}, \bibinfo {author} {\bibfnamefont {G.~Y.}\ \bibnamefont {Chen}}, \
  and\ \bibinfo {author} {\bibfnamefont {Y.~N.}\ \bibnamefont {Chen}},\
  }\bibfield  {title} {\enquote {\bibinfo {title} {Controlling {F}ano resonance
  of nanowire surface plasmons},}\ }\href {\doibase 10.1364/OL.36.003602}
  {\bibfield  {journal} {\bibinfo  {journal} {Opt. Lett.}\ }\textbf {\bibinfo
  {volume} {36}},\ \bibinfo {pages} {3602--3604} (\bibinfo {year}
  {2011}{\natexlab{b}})}\BibitemShut {NoStop}%
\bibitem [{\citenamefont {Chen}\ and\ \citenamefont {Chen}(2012)}]{GYChen_12}%
  \BibitemOpen
  \bibfield  {author} {\bibinfo {author} {\bibfnamefont {G.~Y.}\ \bibnamefont
  {Chen}}\ and\ \bibinfo {author} {\bibfnamefont {Y.~N.}\ \bibnamefont
  {Chen}},\ }\bibfield  {title} {\enquote {\bibinfo {title} {Correspondence
  between entanglement and {F}ano resonance of surface plasmons},}\ }\href
  {\doibase 10.1364/OL.37.004023} {\bibfield  {journal} {\bibinfo  {journal}
  {Opt. Lett.}\ }\textbf {\bibinfo {volume} {37}},\ \bibinfo {pages}
  {4023--4025} (\bibinfo {year} {2012})}\BibitemShut {NoStop}%
\bibitem [{\citenamefont {Chen}(2016)}]{GYChen_16}%
  \BibitemOpen
  \bibfield  {author} {\bibinfo {author} {\bibfnamefont {G.~Y.}\ \bibnamefont
  {Chen}},\ }\bibfield  {title} {\enquote {\bibinfo {title} {Probing the
  spectral density of the surface electromagnetic fields through scattering of
  waveguide photons},}\ }\href {\doibase 10.1038/srep21673} {\bibfield
  {journal} {\bibinfo  {journal} {Sci. Rep.}\ }\textbf {\bibinfo {volume}
  {6}},\ \bibinfo {pages} {21673} (\bibinfo {year} {2016})}\BibitemShut
  {NoStop}%
\bibitem [{\citenamefont {Kuo}\ \emph {et~al.}(2016)\citenamefont {Kuo},
  \citenamefont {Chen},\ and\ \citenamefont {Chen}}]{PCKuo_16}%
  \BibitemOpen
  \bibfield  {author} {\bibinfo {author} {\bibfnamefont {P.~C.}\ \bibnamefont
  {Kuo}}, \bibinfo {author} {\bibfnamefont {G.~Y.}\ \bibnamefont {Chen}}, \
  and\ \bibinfo {author} {\bibfnamefont {Y.~N.}\ \bibnamefont {Chen}},\
  }\bibfield  {title} {\enquote {\bibinfo {title} {Scattering of nanowire
  surface plasmons coupled to quantum dots with azimuthal angle difference},}\
  }\href {\doibase 10.1038/srep37766} {\bibfield  {journal} {\bibinfo
  {journal} {Sci. Rep.}\ }\textbf {\bibinfo {volume} {6}},\ \bibinfo {pages}
  {37766} (\bibinfo {year} {2016})}\BibitemShut {NoStop}%
\bibitem [{\citenamefont {Miroshnichenko}\ \emph {et~al.}(2010)\citenamefont
  {Miroshnichenko}, \citenamefont {Flach},\ and\ \citenamefont
  {Kivshar}}]{Miroshnichenko_10}%
  \BibitemOpen
  \bibfield  {author} {\bibinfo {author} {\bibfnamefont {A.~E.}\ \bibnamefont
  {Miroshnichenko}}, \bibinfo {author} {\bibfnamefont {S.}~\bibnamefont
  {Flach}}, \ and\ \bibinfo {author} {\bibfnamefont {Y.~S.}\ \bibnamefont
  {Kivshar}},\ }\bibfield  {title} {\enquote {\bibinfo {title} {{F}ano
  resonances in nanoscale structures},}\ }\href {\doibase
  10.1103/RevModPhys.82.2257} {\bibfield  {journal} {\bibinfo  {journal} {Rev.
  Mod. Phys.}\ }\textbf {\bibinfo {volume} {82}},\ \bibinfo {pages}
  {2257--2298} (\bibinfo {year} {2010})}\BibitemShut {NoStop}%
\bibitem [{\citenamefont {Chen}\ \emph {et~al.}(2017)\citenamefont {Chen},
  \citenamefont {Lambert}, \citenamefont {Shih}, \citenamefont {Liu},
  \citenamefont {Chen},\ and\ \citenamefont {Nori}}]{GYChen_17}%
  \BibitemOpen
  \bibfield  {author} {\bibinfo {author} {\bibfnamefont {G.~Y.}\ \bibnamefont
  {Chen}}, \bibinfo {author} {\bibfnamefont {N.}~\bibnamefont {Lambert}},
  \bibinfo {author} {\bibfnamefont {Y.~A.}\ \bibnamefont {Shih}}, \bibinfo
  {author} {\bibfnamefont {M.~H}\ \bibnamefont {Liu}}, \bibinfo {author}
  {\bibfnamefont {Y.~N~Chen}\ \bibnamefont {Chen}}, \ and\ \bibinfo {author}
  {\bibfnamefont {F.}~\bibnamefont {Nori}},\ }\bibfield  {title} {\enquote
  {\bibinfo {title} {Plasmonic bio-sensing for the {F}enna-{M}atthews-{O}lson
  complex},}\ }\href {\doibase 10.1038/srep39720 (2017)} {\bibfield  {journal}
  {\bibinfo  {journal} {Sci. Rep.}\ }\textbf {\bibinfo {volume} {7}},\ \bibinfo
  {pages} {39720} (\bibinfo {year} {2017})}\BibitemShut {NoStop}%
\end{thebibliography}
\end{document}